\documentclass[12pt,preprint]{aastex}
\usepackage[numberedappendix]{emulateapj5}
\newcommand{\et}{\mbox{et~al.\ }}
\newcommand{\eg}{\mbox{e.g.,}\ }
\newcommand{\ie}{\mbox{i.e.,}\ }
\newcommand{\kms}{\ifmmode {\rm km\,s}^{-1} \else km\,s$^{-1}$\fi}
\newcommand{\lya}{\ifmmode {\rm Ly\,}\alpha \, \else Ly\,$\alpha$\,\fi}
\newcommand{\hb}{H\,$\beta$}
\newcommand{\ha}{H\,$\alpha$}
\newcommand{\feii}{Fe\,{\sc ii}} 
\newcommand{\oiii}{[O\,{\sc iii}]} 
\newcommand{\civ}{C\,{\sc iv}} 
\newcommand{\ciii}{C\,{\sc iii}]}
\newcommand{\siivoiv}{Si\,{\sc iv}+O\,{\sc iv}]}
\newcommand{\heii}{He\,{\sc ii}}
\newcommand{\lam}{$\lambda$}
\newcommand{\mbh}{$M_{\rm BH}$}
\newcommand{\nMuv}{$nM_{\rm BH,UV}$}
\newcommand{\mbhuv}{$M_{\rm BH,UV}$}
\newcommand{\mhb}{$M_{\rm BH}(\rm H\beta)$}

\newcommand{\mhbse}{$M_{\rm BH}(\rm H\beta,S$-E)}
\newcommand{\mhbrms}{$M_{\rm BH}(\rm H\beta, rms)$}
\newcommand{\mhbmean}{$M_{\rm BH}(\rm H\beta, mean)$}
\newcommand{\lLl}{$\lambda L_{\lambda}$}
\newcommand{\Msol}{\mbox{$M_{\odot}$}}

\newcommand{\hst}{{\it HST}}
\newcommand{\HST}{{\it HST}}
\newcommand{\iue}{{\it IUE}}
\newcommand{\lsim}{\stackrel{\scriptscriptstyle <}{\scriptstyle {}_\sim}}
\newcommand{\gsim}{\stackrel{\scriptscriptstyle >}{\scriptstyle {}_\sim}}

\input epsf

\begin{document}
\submitted{To appear in the Astrophysical Journal, June 1, 2002}
\journalinfo{The Astrophysical Journal, June 1, 2002}

\title{Determining Central Black Hole Masses in Distant Active Galaxies. }

\shorttitle{Black Hole Masses in Distant Active Galaxies.}
\shortauthors{Vestergaard}

\author{M.\ Vestergaard}
\affil{Department of Astronomy, The Ohio State University,
	140 West 18th Avenue, \\ Columbus, OH 43210-1173.
	Email: vester@astronomy.ohio-state.edu
}

\begin{abstract}
An empirical relationship, of particular interest for studies of high redshift
active galactic nuclei (AGNs) and quasars, between the masses of their central 
black-holes and rest-frame ultraviolet (UV) parameters measured in single-epoch 
AGN spectra is presented.  This relationship is calibrated to recently
measured reverberation masses of low-redshift AGNs and quasars.  
An empirical relationship between single-epoch rest-frame optical
spectrophotometric measurements and the central masses is also presented.
The UV relationship allows reasonable estimates of the central masses 
to be made of high-redshift AGNs and quasars for which these masses 
cannot be directly or easily measured by the techniques applicable to 
the lower luminosity, nearby AGNs.  The central mass obtained by this 
method can be estimated to within a factor of $\sim$3 for most objects.
This is reasonable given the intrinsic uncertainty of a factor less than 2 in 
the primary methods used to measure the central masses of nearby inactive 
and active galaxies, namely resolved gas and stellar kinematics in the 
underlying host galaxy and reverberation-mapping techniques.  The UV 
relationship holds good potential for being a powerful tool to study 
black-hole demographics at high redshift as well as to statistically 
study the fundamental properties of AGNs.
The broad line region size $-$ luminosity relationship is key to the
calibrations presented here. The fact that its intrinsic scatter 
is also the main source of uncertainty in the calibrations
stresses the need for better observational constraints to be placed on
this relationship.
The empirically calibrated relationships presented here will be applied 
to quasar samples in forthcoming work.

\end{abstract}


\section{Introduction and Motivation}

The mass of the central black hole, $M_{\rm BH}$, is a
fundamental property of active galactic nuclei (hereafter AGNs) and quasars 
governing the physics of their central engine. 
The mass of the central black hole (or ``massive dark object'') can be measured 
in nearby, inactive galaxies using gas and stellar kinematics (\eg Kormendy \& 
Richstone 1995; Richstone \et 1998; Ho 1999; Kormendy \& Gebhardt 2001) and in 
nearby, active galaxies using results from reverberation mapping studies (\eg 
Koratkar \& Gaskell 1991; Peterson \& Wandel 1999, 2000; Ho 1999; Wandel, 
Peterson, \& Malkan 1999; Kaspi \et 2000).  Recent developments have shown 
the central mass in both inactive and active galaxies to be related to 
properties of its host galaxy, namely the bulge luminosity 
$L_{\rm bulge}$, ({\it inactive galaxies}: \eg Kormendy \& Richstone 1995; 
Magorrian \et 1998; Richstone \et 1998; Ho 1999; {\it active galaxies}: 
\eg Laor 1998; Ho 1999; Wandel 1999; see also Laor 2001; McLure \& Dunlop 2001; 
Wandel 2001) and the bulge stellar velocity dispersion, 
$\sigma$ ({\it inactive galaxies}: 
Ferrarese \& Merritt 2000; Gebhardt \et 2000a; see also Merritt \& 
Ferrarese 2001a; {\it active galaxies}: Gebhardt \et 2000b; Ferrarese \et 2001).  
This is indicative of related formation processes of the central mass and the 
bulge in the host galaxy. The reasons for these purely empirical relationships, 
and that the \mbh{} $-$ $\sigma$ relationship is tighter than 
the \mbh{} $-$ $L_{\rm bulge}$ relationship, are not clear, but several attempts to 
explain them exist (\eg Silk \& Rees 1998; Haehnelt \& Kauffmann 2000; Adams, 
Graff, \& Richstone 2001; Burkert \& Silk 2001; see also \S~5 of Merritt \& 
Ferrarese 2001b). Further study of these relationships and how they may 
change with redshift is one of many motivations for studying
supermassive black holes at high redshift.

Spatially resolved gas and stellar kinematical studies are most useful for 
measuring central masses in inactive and weakly active galaxies. This is because 
in AGNs the strong glare from the central non-stellar continuum source inhibits 
measurements of the stellar absorption features, and the narrow-line gas 
kinematics are perturbed by non-gravitational forces. 
Reverberation (or echo-) mapping techniques (\eg Blandford \& McKee 1982; 
Peterson 2001a) do not require high spatial resolution but instead utilize the 
intrinsic continuum source variability 
and light travel time delays between the location of the ionizing continuum 
source and the line-emitting gas responding to the changing continuum. If the 
line-emitting gas is gravitationally bound to the central black hole,
then simple virial arguments imply that the central mass can be measured.
Reverberation analysis techniques have in recent years been improved to provide
reasonable estimates of the size of the broad line region (BLR) and thereby
to provide reasonable central mass estimates. The accuracy of size 
determinations depend on the data quality, temporal sampling, duration of the 
monitoring campaigns (Collier, Peterson, \& Horne 2001; Polidan \& Peterson 2001), 
and analyses techniques 
(\eg Peterson \et 1998b; Welsh 1999; see also Horne 2001).  Work is in progress 
to reanalyze the available AGN Watch data\footnote{All the data from the AGN 
Watch monitoring campaigns are available at \hfill 
\\ {\tt http://www.astronomy.ohio-state.edu/$\sim$agnwatch} } with better
analysis techniques to improve the central mass estimates for the Seyfert 1 
galaxies which currently have large uncertainties (B.~M.~Peterson 2001, private
communication). Ferrarese \et (2001) discuss the cause of these large errors.
The simple underlying idea of echo mapping is that the measured time delays 
and the behavior and properties of the varying continuum and line emission 
indeed measure distances and velocity dispersion in the BLR and thus measure 
the central mass with reasonable accuracy. This assumes that the black hole 
gravity dominates radiation pressure effects on the BLR and evidence in favor 
of this now exists (Gaskell 1988; Peterson \& Wandel 2000; Gebhardt \et 2000b; 
Ferrarese \et 2001). 
Arguments have been presented that it is to be expected, theoretically, that 
other processes than gravity, such as radiation pressure, anisotropic emission, 
and projection effects, should introduce large systematic uncertainties which 
dominate and prevent accurate measurements of the central masses (Krolik 2001). 
It is worth keeping in mind that theoretical considerations are equally 
affected by our limited knowledge on the geometry and the details of the 
kinematics in the central regions of AGNs. This seems founded by the recent 
studies, which find good consistencies in the central masses of nearby AGNs 
measured both with reverberation and stellar kinematical techniques (Gebhardt 
\et 2000b; Ferrarese \et 2001).

Central masses can now be determined with much smaller uncertainties 
(less than a factor of 2 
depending\footnote{This is judged in part from the uncertainties in the 
masses due to the propagated measurement uncertainties in R$_{\rm BLR}$ 
and $L_{\lambda}$\,(5100\AA) determined in this work (see also 
\S~\ref{errors}) and the reverberation mass uncertainties quoted by 
Peterson \& Wandel (2000) and Ferrarese \et (2001).  Ferrarese \& Merritt 
(2000) quote errors in the stellar kinematics based on the M $-$ $\sigma$ 
relationship to be of order 30\%, but the scatter can be
up to a factor of $\sim$2 (Merritt \& Ferrarese 2001a).} on the method and data quality)
than the early crude estimates could provide (\eg Dibai 1980; Wandel \& 
Yahil 1985; Wandel \& Mushotzky 1986; Padovani \& Rafanelli 1988; 
Padovani, Burg, \& Edelson 1990).
The accuracy of reverberation mass determinations may further improve 
given the continuously improving reverberation analyses techniques.  
These recent developments allow us to revisit with modern techniques 
fundamental issues such as how AGN supermassive black holes 
evolve and relate to galaxy formation and evolution
(\eg Richstone \et 1998; Fabian 1999; Kauffmann \& Haehnelt 2000).
While reverberation techniques and stellar and gas kinematical studies 
will yield valuable insight to the local black-hole demography, 
the situation is different for high-redshift AGNs.
In principle, the \mbh{} $-$ $\sigma$ relationship is sufficiently tight to allow 
estimates of the central AGN mass based on the measured velocity dispersion,
$\sigma$, of the central host galaxy spectrum. This requires, however, a 
significant host galaxy contribution in the AGN spectrum. At the time of 
writing this can only be efficiently studied for low-luminosity, nearby AGNs 
(\eg Nelson \& Whittle 1995; Ferrarese \et 2001).  
Reverberation mapping techniques can in principle be applied to distant AGNs.  
Unfortunately, such studies are also very telescope- and time-consuming (\eg 
Peterson 2001a,b), especially for the more distant and fainter AGNs and for 
the (intrinsically) more luminous quasars, which vary on longer time scales 
(\eg Kaspi 2001). 
The method is also highly dependent on the objects actually showing intrinsic 
continuum luminosity variations when observed.  As a result, obtaining accurate 
reverberation mapping masses of a large, representative sample of the distant 
AGN and quasar populations is close to impossible in a human lifetime.  
Nevertheless, significant advances can be made with reasonable
estimates of the central black hole masses in these objects even to within a
factor of a few. This is particularly useful if the mass estimates are 
based on data that are relatively easy to obtain such as a single-epoch spectrum.
The best available approach is likely the ``ladder'' type of calibration, 
known from the distance scale determination (\eg Freedman \et 2001). That is, 
one calibrates appropriate parameters, more easily obtained for higher 
redshift objects, to the mass measurements at low-$z$. Such an approach is 
pursued here. A relationship between \mbh{}, the \hb{} line width, and the 
optical continuum luminosity, $L_{\lambda}$(5100\AA), is already established 
[see eqns.~(\ref{mrl.eq}) and~(\ref{blrsize.eq}) later]
through the confirmation of the theoretically expected relationship between 
the BLR size, $R_{\rm BLR}$, and $L_{\lambda}$(5100\AA) [Wandel \et 1999; 
Kaspi \et 2000]. 
The time delay, $\tau$, and hence $R_{\rm BLR} = c\tau$ for \civ \,\lam\,1549 
is about half that of \hb{} (Korista \et 1995), and multiple broad emission 
lines in NGC\,5548 exhibit the same virial relationship (Peterson \& Wandel 
1999, 2000), as expected if reverberation techniques work. 
Therefore, it is reasonable to assume that a similar relationship can be 
obtained with appropriate rest-UV spectral measurements. 
Such a relationship is of particular interest if single-epoch 
spectra can supply the required rest-frame UV measurements.  
Then it becomes straightforward to estimate central masses 
for large samples of high-redshift AGNs. If uncertainties in such 
an approach are, relatively speaking, reasonable this is a powerful tool 
for statistical studies and black hole demography at high redshift. 
This is the focus of this paper.

The \civ \,\lam 1549 emission line is chosen here to provide a 
characteristic velocity dispersion appropriate for the calibration of 
single-epoch UV spectral measurements for several reasons.  First,
this line is accessible from the ground for objects with redshifts
between $\sim$1 and $\sim$5. Second, its profile is not commonly affected
by strong or numerous absorption lines typical of the \lya{} emission 
line\footnote{The exceptions are broad absorption troughs in the blue 
profile wing seen in a small fraction of quasars. Narrow associated 
absorption lines, detected in some quasars, are generally easily 
corrected for.}.
Also, its line width is also much less affected by blending effects from
other line emission, including \feii{} emission, than either of the 
\ciii \,\lam 1909 and \siivoiv \,\lam 1400 lines. Contaminating \feii{} 
and \heii \,\lam 1640 emission mainly affect the lower profile wings of 
\civ{} (\eg Figure~2 by Marziani \et 1996; Vestergaard \& Wilkes 2001; 
M. Vestergaard et al., in preparation).
For these reasons it is also an emission line on which future rest-frame
UV and optical reverberation mapping analyses may be applied in both 
nearby and more distant AGNs, providing tests of the calibrations 
presented here.

It is important to point out that such calibrations do not eliminate the 
need for further reverberation mapping studies. Rather, they emphasize the 
strong need to obtain yet better and more monitoring data. 
First, this work and the BLR size -- luminosity relationship are based on 
samples of Seyferts and quasars that are, after all, small ($\lsim$34 
objects) and do not span the full range of observed AGN properties. 
Second, the UV and optical calibrations presented in this work depend 
strongly on the size $-$ luminosity relationship and the intrinsic scatter 
around this relationship is the main source of  uncertainty in the 
calibrations (\S~\ref{optcal} and \S~\ref{uvcal}).  The presence of this 
scatter thus accentuates the need for more and better BLR size determinations. 
It is now quite clear that significant improvements thereof require and can
be obtained by a fully dedicated set of space-based telescopes, which can 
provide sufficient temporal sampling and duration of the monitoring campaigns
(Collier \et 2001), and 
simultaneous observations across multiple wavelength bands (Peterson 2001b;
Netzer 2001) to constrain the transfer functions and the BLR geometry better.
In return, more accurate measurements can be obtained of the BLR size, 
and hence of the central masses (\eg Horne \et 2002).  This is just one 
of the important issues that the proposed MIDEX mission, {\it Kronos}, 
will address if selected.  Given the practical difficulties of long 
term multi-wavelength monitoring with existing facilities on the ground and 
in space (\hst{} and X-ray telescopes), including obtaining bona-fide 
simultaneous observing time on many different telescopes for long periods 
of time, {\it Kronos} may quite possibly be our best opportunity to study 
how the central regions of nearby and distant AGNs are structured and 
``engineered'' (Blandford 2001).  Ultimately, that will allow more accurate 
mass estimates for distant AGNs than are possible at present, as discussed
above.

This work was inspired by the recently established BLR size $-$ luminosity
relationship (Kaspi \et 2000), which is applicable not only for nearby
Seyfert 1 galaxies but also for the intrinsically brighter quasars. 
Hence, assuming this size $-$ luminosity relationship applies to more
distant quasars also, their central masses can thus be estimated.
Here, relationships between \mbh{} and single-epoch rest-frame optical 
and UV measurements, respectively, are derived and the uncertainties 
introduced by this approach are evaluated. 
It will be shown that single-epoch spectrophotometry can be used to 
estimate the central masses to within a factor of 3 with high probability,
rendering the calibrations very useful.

In what follows `optical' and `UV' refer to these same bands in the 
{\it rest-frame} of the AGN or quasar.  A cosmology with 
H$_0$ = 75 ${\rm km~ s^{-1} Mpc^{-1}}$, q$_0$ = 0.5, and $\Lambda$ = 0 
is used throughout.  

The paper is structured as follows.
The method adopted for the calibration is outlined in \S~\ref{method}.
Section~\ref{data} describes the data considered, \S~\ref{secal} 
discusses how well representative the single-epoch optical data are of the 
multi-epoch reverberation data, \S\S~\ref{optcal} and~\ref{uvcal} 
present calibrations of single-epoch optical and UV measurements, 
respectively.  In \S~\ref{errors} the performance of 
the adopted calibrations is evaluated. 

\vskip 1cm

\section{The Method \label{method}}

The virial theorem is used to estimate the mass of the central black hole, 
$M_{\rm BH} \approx r v^2/G$, where $v$ is the velocity dispersion of matter 
at distance, $r$, which is gravitationally bound to the black hole. 
The velocity dispersion can be expressed as $v = f\,\cdot v_{\rm FWHM}$ where
$v_{\rm FWHM}$ is the FWHM of the emission profile of the broad line gas.
The factor $f$ is on the order of unity and depends on the geometry and the details 
of the kinematics (\eg Peterson \& Wandel 1999, 2000; Fromerth \& Melia 2000;
Krolik 2001; McLure \& Dunlop 2001). The central mass can be expressed as:

\begin{equation}
M = 1.5 \times 10^5 \left(\frac{R_{\rm BLR}}{\rm lt-days}\right)
\left(\frac{v_{\rm FWHM}}{10^3~ \rm km~s^{-1}}\right)^2 \Msol
\label{mrl.eq}
\end{equation}
\vskip 0.5cm
Based on 17 nearby Seyfert~1 galaxies (Wandel \et 1999) and 17 Palomar-Green 
(hereafter PG; Schmidt \& Green 1983; Green, Schmidt, \& Liebert 1986)
quasars, Kaspi \et (2000) determine an empirical relationship
between the size of the broad line region, $R_{\rm BLR}$, and the continuum 
luminosity, $\lambda L_{\lambda}$(5100\AA), where $R_{\rm BLR}$ is the distance 
of the emission-line clouds responding to the central continuum variations as
determined from reverberation studies [eq.\ (6) of Kaspi et al. 2000]:

\begin{equation}
R_{\rm BLR} = (32.9^{+2.0}_{-1.9})~\left[ \frac{ \lambda L_{\lambda}
\rm (5100\AA) }{ 10^{44}~\rm ~ergs\,s^{-1}} \right]^{0.700\,\pm\,0.033}~~{\rm 
lt-days}
\label{blrsize.eq}
\end{equation}

In appendix~\ref{sizelum} this relationship is re-derived using regression 
analysis appropriate for data with internal scatter. A critical assessment of
the object sample is also performed. The modified luminosity dependence has a
larger uncertainty, which is probably more realistic given the intrinsic scatter, 
than that quoted above. However, within this uncertainty the approach adopted 
in appendix~\ref{sizelum} does not yield an $R_{\rm BLR} - L$ relationship 
significantly different from that of Kaspi et\,al.\ for the current sample of 
AGNs; this may change once more data are available to better constrain the 
$R_{\rm BLR} - L$ relationship.
Therefore, in what follows eqn.~(\ref{blrsize.eq}) is used to estimate the 
BLR size when unknown. 

Equation~(\ref{mrl.eq}) assumes that the broad line emitting gas is 
gravitationally bound, and that the cloud velocity dispersion is 
isotropic such that $v \,=\, \sqrt{3} \,|\sigma_i|, ~i\,=\,1,2,3~;~|\sigma_i| 
\,=\, v_{\rm FWHM}$/2; $v_{\rm FWHM}$ is in units of \kms.
The details of the geometry and kinematics, \ie the exact value of $f$, are 
unknown but have been debated (\eg Peterson \& Wandel 1999; Fromerth \& Melia
2000; McLure \& Dunlop 2001).  
Most expected values of $f$ are anticipated to result in a constant 
offset in $\log$ \mbh{} (Krolik 2001). Because such details are not yet 
clarified or well constrained (\eg Wandel 2001) the approach taken in this 
work is to compute the central masses with basic and common assumptions
to allow comparison with most other work. 
The offsets in mass estimates due to different assumptions of geometry and 
kinematics should however be kept in mind when considering the \mbh{} values 
in absolute terms.  

Assuming that equation~(\ref{blrsize.eq}) is valid for all active 
galaxies, we therefore have an approximative relationship between 
the mass of the central black hole, the AGN continuum luminosity at 5100\AA{} and 
the line width of the H$\beta$ emission-line [eqn.~(\ref{mrl.eq})].
Given that the BLR size $-$ luminosity relationship is determined from
line-continuum variability of the objects, the specific \hb{} line width to be 
used to 
determine the `reverberation mass' is that appropriate for the emission-line gas 
at the distance $R_{\rm BLR}$ from the ionizing continuum source that is 
varying. That is, the FWHM of \hb{} in the `root-mean-square\footnote{Based 
on the spectral database obtained during a monitoring campaign a `mean
spectrum' can be generated, which is the mean flux at a given wavelength
bin of all the individual spectra. Similarly, the root-mean-square deviation
at each wavelength bin from this mean can be obtained. This rms spectrum
shows how the object spectrum varies in response to continuum emission
variations. See Peterson \et (1998a) for details and examples of `rms' and 
`mean' profiles. }
(rms) spectrum' (e.g., Peterson \& Wandel 1999, 2000) will be used below for 
the variability data.
Kaspi \et (2000) argue that FWHM(\hb, mean), the FWHM of \hb{} in the 
`mean spectrum', may equally well be used. In the interest of completeness and 
for comparison, the masses based on the mean optical spectra are included in 
the analysis below, even if the rms masses are considered, strictly speaking, 
the most representative of the intrinsic, actual central masses.

The method used here to calibrate single-epoch UV measurements to reflect the 
central black-hole masses is briefly as follows. 
First, the single-epoch FWHM(\hb) and continuum luminosities are compared 
with the multi-epoch equivalents to test how representative the single-epoch
measurements are, and whether systematic offsets between these two sets of 
measurements exist (\S~\ref{secal}). If that is the case, a calibration 
or modification of the single-epoch measurements may be appropriate to ensure 
they introduce minimum systematic error or do not add unnecessary scatter.
Then, the single-epoch mass estimates based on optical measurements, \mhb, 
are computed using equations~(\ref{mrl.eq}) and (\ref{blrsize.eq}). These 
estimates are compared and calibrated to the available reverberation masses, 
\mhbrms, based on \hb{} measurements for the PG quasars (\S~\ref{optcal}). 
The next step is to determine the best calibration of an appropriate combination
of the single-epoch FWHM(\civ) and $\lambda$ $L_{\lambda}$(1350\,\AA) 
measurements to the best available \mbh{} measurements (\S~\ref{uvcal}). 
This is done for the subset of 26 objects with both \mhbrms{} and UV measurements 
available.  How well this method yields representative central mass estimates
is discussed in \S~\ref{errors}.

\section{Data \label{data}}

The calibration of optical single-epoch spectral measurements is based on
the 19 objects common to the sample of Seyfert\,1s and PG quasars with 
established reverberation masses (Peterson \& Wandel 1999, 2000; 
Wandel \et 1999; Kaspi \et 2000) and the 
PG quasars studied by Boroson \& Green (1992; hereafter BG92). 
The calibration of the single-epoch UV measurements is based on the sample
(labeled ``UVrev'') of 26 reverberation AGNs and PG quasars (Wandel \et 1999; Kaspi \et 
2000) with available UV spectroscopy. Mostly for comparison, a sample is also 
considered consisting of UVrev and 30 additional PG quasars (BG92) for which 
\civ{} line widths and UV continuum luminosities are also available in the 
literature. These data are described in the following.  

\subsection{Optical Data \label{OptData}}

BG92 list H$\beta$ FWHM measurements of single-epoch spectra of many quasars 
in the PG sample from which the contaminating iron emission and the narrow 
line contribution are subtracted; i.e. the FWHM is that of the supposedly 
intrinsically emitted {\it broad} H$\beta$ line emission. Some of the
line widths were substituted with measurements obtained in this work, because
the subtraction of the narrow component renders the profile less
representative of the line width needed for this study (\S~\ref{FWcompar}).

Monochromatic luminosities, $L_{\lambda}$\,(5100\AA), are determined for
most of the PG quasars from the spectrophotometry by Neugebauer \et (1987).
The flux densities at 5100\AA{} are approximated by a linear interpolation.
For most objects this interpolation is done over short or neighboring 
frequency ranges in relatively tight spectral energy distributions, and so 
should be reliable.
For the few sources not included in this database $L_{\lambda}$(5100\AA) is 
computed based on the 4400\,\AA{} flux densities listed by Kellerman \et (1989).
These are based on B-band photometry of Schmidt \& Green 
(1983).  A power-law continuum, F$_{\nu} \propto \nu^{\alpha}$, with an 
average optical slope, $\alpha ~=~ \alpha_{\rm opt}$ = $-$0.5  is assumed and 
extrapolated to 5100\,\AA{}.  
The optical data are listed in Table~\ref{Opt_pars.tab}. 

Errors on the parameters are propagated using the following assumed 
uncertainties: (1) spectral slopes, $\sigma (\alpha)$=0.2, reflecting the 
different slopes commonly adopted in the literature for quasars (\eg 
Francis \et 1991: $\alpha$=$-$0.3; V\'eron-Cetty \& V\'eron 1993: 
$\alpha$=$-$0.7), (2) redshift, $\sigma (z)$=0.0025, a compromise between 
the typical uncertainty for redshift determinations based on optical and 
low-ionization lines, and those relying on UV high-ionization lines only 
where velocity shifts may exist between lines, especially for higher 
redshift quasars [$\sigma(z)$ has a negligible effect], (3) Neugebauer 
\et (1987) quote flux errors for each measurement and for each object, 
and (4) the B-magnitude uncertainty is 0.27 mag (Schmidt \& Green 1983).
BG92 do not quote the uncertainty on their FWHM(\hb) measurements.  
Based on the few error quotes available in the literature (e.g.  
Brotherton 1996; Vestergaard 2000; M. Vestergaard \et, in preparation) a 
reasonable relative error on the FWHM measurements of broad emission lines is 
$\sim$10\% depending on the measurement method and the quality of the data. 
Blending effects will increase this uncertainty.  The uncertainties in the 
BLR size $-$ luminosity relationship [eqn.~(\ref{blrsize.eq})] are also taken 
into account; see also Appendix~\ref{sizelum}. 

\subsection{UV Data}

The single-epoch UV measurements were obtained from a handful of studies 
including Wilkes \et (1999), Wang, Lu, \& Zhou (1998), Laor \et (1994, 1995), 
Koratkar \& Gaskell (1991), and Wills \et (1995).  Line widths and 
luminosities were generally taken from the same study.  For a few objects 
with data only presented by Marziani \et (1996) the line widths were remeasured 
directly on the published \civ{} profiles. This was done because the authors 
decompose the profiles and only list measurements for the individual broad 
and narrow components. As argued in \S~\ref{FWcompar} the line widths 
for this study should be measured on profiles including the narrow component as 
it originates in the BLR like the broad component; this is contrary to the 
assumption of Marziani \et (1996).  Most of these earlier studies present UV 
continuum luminosities at 1350\AA{} or quite close to it. Wilkes \et (1999) 
quote continuum luminosities at 2500\,\AA, extrapolated from available V and B 
band photometry.  Most of the objects studied by Wilkes \et were also studied 
by others who do measure the continuum luminosities directly. These latter 
measurements are less subject to extrapolation uncertainties and were thus 
preferred.
All the luminosities adopted from the literature were extrapolated to be 
a measure of the 1350\,\AA{} luminosity, if necessary, assuming the slope 
and cosmology quoted above; this extrapolation is in all cases over a 
relatively short wavelength range ($\lsim$100\,\AA).  For the measurements 
with quoted uncertainties the errors were propagated, as outlined above. 
For line widths with no measurement errors a conservative uncertainty of 
10\% was imposed. Typical luminosity uncertainties of 30\% were
adopted for luminosities without quoted errors. The discussion in 
\S~\ref{Lcompar} show this is reasonable.

For some of the objects two or more studies quote line widths and luminosities.
It was then assumed that the differing measurements is a consequence of
intrinsic scatter due both to variability and to measurement uncertainty. 
The mean of the measurements was therefore adopted and the difference in
line width was adopted as the uncertainty in the FWHM measurement, unless it 
was smaller than the adopted conservative 10\% error. The luminosities
were generally rather similar.  The adopted UV measurements and 
their source references are listed in Table~\ref{UV_pars.tab}.

\section{Calibration of Single-Epoch Optical Measurements \label{secal}}

Before the single-epoch spectral measurements (line width and 
monochromatic luminosity) are calibrated to the (multi-epoch) reverberation 
measurements it is important to compare these two types of measurements 
to examine whether single-epoch measurements 
are reasonably representative of the reverberation measurements, to what 
degree and whether certain spectral corrections are necessary. 
This includes an attempt to understand how and in what way they may differ. 
This direct comparison is performed on the objects common to the Kaspi \et 
(2000) and BG92 studies, namely the 17 PG quasars studied by Kaspi \et and
two Seyfert galaxies, Mrk\,110 and Mrk\,335, (Wandel \et 1999), all of 
which were studied with reverberation mapping techniques.

\placefigure{FWHM.fig}

\subsection{Line Width Measurements \label{FWcompar}}

The appropriate BLR velocity dispersion to use in the mass estimate 
is that measured from the `rms' profile (\S~\ref{method}).
Because the mass depends strongly on the line width [eqn.~\ref{mrl.eq}]
it is important to identify the single-epoch line profile that 
best represents the rms profile. 
That is, is it important to subtract contaminating 
\feii{} emission, and/or perhaps the narrow line contribution before 
measuring the FWHM? Another goal of this section is to quantify how
well a single-epoch line width measurement estimates the rms line width.
Rms spectra show the variable part of the continuum and line emission 
and its strength as a function of wavelength. 
The \hb{} line width in the rms spectrum, hereafter FWHM(\hb, rms), is 
the width of the variable part of the line.  Rms spectra can have 
different line emission contribution from single-epoch spectra
(examples of mean and rms profiles are presented by Peterson \et 1998a).
The \hb{} line width measured in a single-epoch spectrum, hereafter 
FWHM(\hb), is expected to be more similar to the mean of many individual
(single-epoch) spectra than FWHM(\hb, rms), which depends on which part
of the broad line gas that varies. 
Therefore, in what follows, after a general discussion of
the single-epoch and multi-epoch line widths, the FWHM(\hb) values are first
discussed in relation to FWHM(\hb, mean) and then to FWHM(\hb, rms).

\subsubsection{The Best Single-Epoch Profile to Measure}

Figures~\ref{FWHM.fig}a and~\ref{FWHM.fig}b show the FWHM(\hb, mean) and 
FWHM(\hb, rms), respectively, and their uncertainties measured by Kaspi \et{} 
and Wandel \et{} in their multi-epoch spectra are plotted for the 18 objects 
of those mentioned above with \hb{} measurements versus the single-epoch FWHM(\hb) 
measurements by BG92.  Note that there is a typographical error in 
Table~2 of BG92, FWHM(\hb) = 5320 \kms{} for PG1307$+$085 (Laor 2000).  
The dotted line represents a one-to-one relationship. 
With a few exceptions (labeled in Fig.~\ref{FWHM.fig}) the single-epoch
FWHM(\hb) shows a general consistency with FWHM(\hb, mean) within 15\% $-$
20\% variation in the single-epoch width and also show consistency with 
FWHM(\hb, rms) to within $\sim$20\% $-$ 25\% variation. This is consistent with 
the $\pm$15\% line width variation (B.\ M.\ Peterson 2001, private communication)
observed for NGC\,5548 during 1988 December 14 $-$ 1996 October 16 (\eg 
Peterson \et 1999), where the \hb{} line flux varied up to $\pm$66\%.  It 
indicates that 
most of the single-epoch FWHM(\hb) measurements and their deviation from the 
multi-epoch measurements can perhaps be accounted for as being due to intrinsic 
continuum and line variations. A detailed discussion of these deviations
follows below.

It is important to note that there are technical differences between the 
BG92 and the multi-epoch measurements.  BG92 fit and subtract the optical 
\feii{} emission around \hb{} and \oiii \,\lam \lam 4959,\,5007 and also 
fit and subtract the narrow line \hb{} contribution. 
Kaspi \et and Wandel \et do not perform any of these corrections to their 
data.  Subtraction of the narrow line contribution will increase the FWHM 
value.  This increase may be quite significant depending on the strength 
of this contribution (see below and \eg Jackson \& Browne 1991). 
By eliminating the \feii{} emission, which blends into the red wing 
of \hb{} and both of the \oiii{} doublet lines, a smaller \hb{} width is
obtained. This holds as long as the contaminating \feii{} emission does not 
artificially increase the underlying continuum level significantly, creating 
a so-called ``pseudo-continuum'' (\eg Wills, Netzer, \& Wills 1985; 
Vestergaard \& Wilkes 2001).
Subtracting the \feii{} emission will then tend to increase the 
FWHM as the line peak height is increased (the local continuum level or 
``zero-point'' of the line is lowered).  Thus, the effects of subtracting the 
\feii{} emission is not straightforward to predict a priori. 

For this type of study the FWHM(\hb) measurement that best represents 
FWHM(\hb, rms)
is one measured on a profile for which {\it only} a correction for strong \feii{} 
contamination is performed, and especially one in which the narrow line contribution 
is {\it not} subtracted. This is justified in the following.
Note that if the \feii{} emission is essentially constant in strength while
the object was monitored, the \feii{} emission contribution is automatically
eliminated in the rms spectra. AGN optical \feii{} emission does not seem
to vary much or very fast if at all (\eg Wamsteker \et 1990; Kollatschny \& 
Welsh 2001) so it is 
reasonable to assume that the FWHM(\hb, rms) is not significantly affected 
by \feii{} emission, if present.
For this reason it is desirable to use single-epoch FWHM(\hb) measurements
which are not affected by the presence of \feii{} emission.

For these reasons and because some of the BG92 FWHM(\hb) measurements deviate
strongly from the FWHM(\hb,rms) of Kaspi \et and Wandel \et (see below and 
Fig.~\ref{FWHM.fig}), 
the line width was remeasured in this work in the {\it original} spectra (\ie 
those {\it not} corrected for \feii{} or narrow component emission; these data 
were kindly provided by T. Boroson) of the PG quasars studied by Kaspi \et 
With exception of a handful of objects with strong \feii{} emission and/or 
a strong narrow line emission component, these non-corrected FWHM values are 
consistent with the BG92 values to within the (assumed) 10\% measurement 
uncertainties; the inconsistent measurements were corrected as described below.  
The objects showing ``uncorrected'' FWHM(\hb) values larger than the BG92 
measurements are all \feii -strong, indicating that the \feii{} blending effects 
in the red wing of \hb{} dominate the effects of the alleged \feii 
-pseudo-continuum for these objects.  
For these objects the \feii{} corrected BG92 measurements were chosen to be the 
most representative of FWHM(\hb, rms) since this width is not significantly
affected by \feii{} emission.
However, the uncertainties of the BG92 FWHM(\hb) measurements for these objects 
were corrected to reflect the effect of the \feii{} emission. The error was set 
to the difference between the measurements before and after the \feii{} emission 
(and narrow component) subtraction if this difference was larger than the 
conservatively assigned 10\% error. The most strongly \feii{} contaminated quasar, 
PG1700$+$518 show an FWHM of $\sim$650 \kms{} larger than the BG92 value 
(Fig.~\ref{FWHM.fig}).

PG1704$+$608 clearly illustrates why the narrow line core should not be subtracted.
PG1704$+$608  has a very strong narrow component and exhibits a highly significant 
offset in FWHM(\hb) from FWHM(\hb, mean). 
Its rms spectrum shows that the strongest \hb{} variation furthermore occur in 
the narrow line core [the FWHM(\hb, rms) is quite \mbox{small;} Figure~\ref{FWHM.fig}b], 
explaining why the BG92 FWHM is far from representative of the typical 
FWHM(\hb, mean) and FWHM(\hb, rms) measured for this object. This means
that the low-velocity gas emitting \hb{} is indeed well within the BLR and that 
the narrow line core is {\it bona-fide} BLR line emission.  Typical variability 
time scales of the narrow emission component support this (Stirpe 1990).  
For PG1704$+$608 the BG92 FWHM(\hb) was therefore replaced, in the analysis 
described below, by the FWHM(\hb) measured here in the uncorrected spectrum.  

\subsubsection{Line Width Deviations \label{FWdeviation}}

The objects labeled in Figure~\ref{FWHM.fig}a (except PG1617$+$175) have BG92
FWHM(\hb) measurements larger than FWHM(\hb, mean) in excess of $\sim$20\%
variation and the typically expected measurement uncertainties.
Are these deviating measurements easily understood?
PG0052$+$250 and PG1307$+$085 are quite variable in the \hb{} line flux as 
indicated by the light curves of Kaspi \et (2000) and so their deviating line 
widths are not unexpected.
The ``uncorrected'' FWHM(\hb) of Mrk\,110 and PG1426$+$015 deviate from
the BG92 measurements by more than 10\%. Both objects have strong
(spiky) narrow \hb{} components (see e.g., Fig.~1 by BG92)
that were subtracted by BG92. This explains the larger BG92 widths.
PG1613$+$658 behaves strangely. The mean \hb{} profile is broad but it
varies 
most strongly in the narrow line core (see \eg Figure~\ref{FWHM.fig}b
and Table~6 and Figure~5 of Kaspi \et 2000).  However, for this object the 
narrow component of neither the single-epoch profile nor the mean profile 
is a fair representation of the rms profile (Figure~\ref{FWHM.fig}b).  
Therefore, this object will commonly be excluded from the analysis. 
To reiterate, for Mrk\,110, PG1426$+$015, and PG1704$+$608 the single-epoch 
FWHMs based on the original, uncorrected data are used in the analysis 
instead of the BG92 values.  For the remaining objects the single-epoch
FWHM(\hb) is consistent with FWHM(\hb, mean)
to within reasonable effects of line variability.

Figure~\ref{FWHM.fig}b shows the single-epoch FWHM(\hb) of BG92 with the
corrections discussed above plotted versus the FWHM(\hb, rms) 
 [Wandel \et 1999; Kaspi \et 2000].
Most of the objects with single-epoch FWHM(\hb) deviating from FWHM(\hb, rms)
by more than $+$15\% were well within $\pm$15\% from the FWHM(\hb, mean) in 
Figure~\ref{FWHM.fig}a.  
This may indicate that the variation in FWHM is typically a little larger 
than 15\%, as is seen for NGC\,5548, in these objects and mostly occur in 
the narrow line core.  The main technical difference between the BG92 
FWHM(\hb) and the FWHM(\hb, rms) is that the BG92 FWHM only reflects the 
width of the broad line component as both line widths are expected to be 
little affected by the presence of \feii{} emission. For objects with 
strong variation in the low-velocity gas the two FWHM measurements will 
therefore differ significantly as observed and discussed above.
Of the two objects (PG1617$+$175 and PG1613$+$658) with single-epoch FWHM(\hb)
(and error) deviating by $\gsim$15\% from FWHM(\hb, rms),
only PG1613$+$658 has such a significantly deviant BG92 FWHM(\hb) that the offset
cannot be ascribed to larger measurement uncertainties or stronger
source variation than the $\pm$15\% shown by NGC\,5548.

It is worth noting for completeness that the differences in spectral resolution
between the BG92 and the Kaspi \et spectra do not contribute to the line width 
differences. The original BG92 spectra were 
degraded to the resolution of $\sim$10 \AA{} of the Kaspi \et spectra for this
analysis.  No significant FWHM differences were found.

\subsubsection{Regression Analysis}

Regression analyses were performed  to statistically test how well the 
single-epoch \mbox{FWHM(\hb)} represents the multi-epoch line widths.  It is 
clear from Figure~\ref{FWHM.fig}a that the objects with line widths below 
4000 \kms{} yield very consistent measurements in spite of their variability, 
while the broader lined objects have larger uncertainties from both
measurements [recall, that a 10\% measurement error was assumed for 
the single-epoch FWHM(\hb)] and especially from intrinsic source 
variability; these objects have rather similar luminosities to the 
other objects in the sample, and so these differences are not likely 
luminosity-related.  However, PG1613$+$658, PG1426$+$015, and PG1617$+$175 
have the broadest mean \hb{} line and also strong, broad \feii{} emission 
(see BG92).  This emphasizes the need for linear regression methods, which 
are more appropriate for the nature of these data.  The regression analyses 
were therefore performed using the bivariate correlated errors and 
intrinsic scatter (hereafter BCES) algorithm (Akritas \& Bershady 1996).  
When intercomparing the multi-epoch measurements with those of 
single-epoch spectra, intrinsic scatter and measurement errors are 
expected. The BCES algorithm is the most appropriate to use as it does 
not, as do many other linear regression methods, assume a perfect 
relationship between the 
variables if the measurement errors could be made insignificantly small. 
Merritt \& Ferrarese (2001a) compare the BCES algorithm to other commonly 
used linear regression methods. 

The uncertainties used in the regressions are the symmetric errors on 
the logarithms based on the positive linear errors.  No significant 
differences were found when using the negative linear errors instead.
These symmetric errors [= $(\log{e})\,(\sigma_X/X)$] were 
determined by propagating the linear errors to the logarithmic 
relationship.

The best fit to the FWHM(\hb, mean) and FWHM(\hb) distribution is 
shown in Table~\ref{regression}.  It is based on sample B, that is, the 
PG quasars (Kaspi \et 2000) and Mrk\,110 and Mrk\,335 (Wandel \et 1999), 
excluding PG1351$+$640 as it has no \hb{} measurements available. 
The BCES regression is quite robust; bootstrapping simulations reproduce
the theoretically expected results well (see e.g., Akritas \& Bershady 
1996).  Also, there is little difference between the BCES(Y$|$X) 
[i.e., Y = f(X)] and BCES(X$|$Y) [i.e., X = g(Y)] regressions.  
The bisector bisects these two regressions and so the said small 
difference is reflected in the similar results obtained for the 
BCES(Y$|$X) and bisector regressions (both are listed in 
Table~\ref{regression}).  These BCES 
regressions for all the data points in Figure~\ref{FWHM.fig}a (sample B) 
are consistent with a unity relationship within 1.8$\sigma$. When the 
FWHM(\hb, rms) measurements are considered (Fig.~\ref{FWHM.fig}b) 
PG1613$+$658 is very much an outlier and it is excluded from the 
regressions to those data as it behaves in an obviously strange fashion
(\S~\ref{FWdeviation}).
For comparison with the results of that analysis, the regressions were 
repeated on the sample in Figure~\ref{FWHM.fig}a excluding this object (sample 
C): the data now display a tighter relationship and the BCES regression is 
consistent with a unity relationship to within 1$\sigma$, with no regard to 
which is the independent variable (Table~\ref{regression}). Only the BCES 
bisector for sample C is shown in Figure~\ref{FWHM.fig}a for visibility.
The BCES bisector is also the most representative of the intrinsic 
relationship between the two line widths, because they are different
measures of the same property.

In Figure~\ref{FWHM.fig}b the bisector regression (solid line) for sample C 
shows consistency with a unity relationship
to within 1\,$\sigma$: FWHM(\hb, rms) = (0.98$\pm$0.08)$\times$ 
single-epoch FWHM(\hb) $-$ (265$\pm$227).  Whether this relationship or 
a pure 1:1 relationship is used, the propagated effect on $\log$\,\mbh{} 
is only of order $\sim$0.04\,dex for a 4000 \kms{} line width (and ranging 
between 0.07 and 0.03 dex for widths of 2000 \kms{} and 6000 \kms, 
respectively) which is well within typical FWHM measurement 
uncertainties. So, assuming a 1:1 relationship is fair, as long as the 
single-epoch line width is measured as described in \S~\ref{FWdeviation}.

\subsubsection{Is the High-Velocity BLR Gas Optically Thin?}

It is interesting to note that this comparison of FWHM(\hb, rms) and 
single-epoch FWHM(\hb) (with or without the \feii{} emission) shows that for 
the PG quasars the single-epoch FWHM(\hb) is predominantly larger than the 
FWHM(\hb, rms).  If all the line flux, both at high and low velocity, in the 
profile is varying with similar amplitudes one would na\"{\i}vely expect the 
rms profile to be consistently broader than any given single-epoch profile as 
the rms profile represents the responding amplitudes and the velocities of the 
responding gas.  
At the very least, the single-epoch width should scatter around FWHM(\hb, rms).
Yet the opposite is seen here. One may speculate whether this indicates that the 
broad line wings, which are often not represented in the rms line profiles, are
mostly due to optically thin material, which does not respond to continuum 
variations as does optically thick gas? 
This is probably seen for \heii{} in NGC\,4051 (Peterson \et 2000).
PG1704$+$608 and, especially, PG1613$+$658 were pointed out above as extreme 
cases showing
broad average profiles, yet exhibit variability in the narrow line core only. 
Other objects showing similar behavior in their \hb\ line profiles are Mrk\,590
(Ferland, Korista, \& Peterson 1990) and Mrk\,335 (Kassebaum \et 1997).
Similar behavior has been seen for both the \lya{} and \civ{} line profiles in
the higher luminosity quasars (\eg O'Brien, Zheng, \& Wilson 1989; P\'erez, 
Penston, \& Moles 1989; Gondhalekar 1990). Therefore, the presence
of high-velocity, optically thin line emission is likely rather common in AGNs
and quasars. Shields, Ferland, \& Peterson (1995) discuss this issue and its
possible importance. They further argue that the presence of such optically
thin emission gas can explain some of the variability properties of Seyfert~1s.

To summarize \S~\ref{FWcompar}, 
statistically, the single-epoch \hb{} line width displays a unity 
relationship with both FWHM(\hb, mean) and FWHM(\hb, rms) to within 
1$\sigma$. However, this is valid as long as the single-epoch FWHM(\hb)
is measured on \hb{} profiles, which {\it includes} the narrow component
but for which especially strong (and blending) \feii{} emission is 
subtracted.

\subsection{Luminosity Measurements \label{Lcompar}}

The optical continuum luminosities for the single-epoch spectra are 
gathered mostly from Neugebauer \et (1987) supplemented with data 
from Schmidt \& Green (1983) for Mrk\,110 and Mrk\,335, as described 
in \S~\ref{data}.  
In Figure~\ref{LumLum.fig} these $\lambda L_{\lambda}$(5100\AA) measurements 
are compared to those determined by Wandel \et (1999) and Kaspi \et (2000) 
for the same sample of quasars discussed above in \S~\ref{FWcompar} but 
including also PG1351$+$640 (i.e, sample A; Table~\ref{regression}); the 
luminosities are plotted with the same cosmology 
(H$_0$ = 75 ${\rm km~ s^{-1} Mpc^{-1}}$, q$_0$ = 0.5, and $\Lambda$ = 0). 
The uncertainties 
listed by Kaspi et al., based on the rms in the continuum light curves, are 
also plotted.  The single-epoch luminosity uncertainty reflects the 
propagated errors based on the measurement uncertainties as described in
\S~\ref{OptData}.
It is clear from Figure~\ref{LumLum.fig} that for most of 
the objects the luminosity measured by Neugebauer \et is stronger. 

Regression analyses show (Table~\ref{regression}) a best fit slope consistent 
with 1.0 and an intercept of zero to within the uncertainties.  
The various BCES results are rather robust and show little 
difference.  Figure~\ref{LumLum.fig} indicates that most of the data 
points are scattered around a slope of one with a systematic offset of 
$\sim$0.1 dex, (i.e., $\sim$30\%).  This offset is equivalent to an offset 
in $\log$\,M of 0.07 dex. As will be clear later, this is well within the 
uncertainties of the (calibrated) mass estimates and will thus not 
significantly affect the calibration of optical single-epoch mass 
estimates.  Therefore the central mass estimate calibration is evaluated 
assuming no luminosity offset, i.e., no correction of either luminosity 
measurements is applied, but keeping the offset in mind.
The $\sim$0.1\,dex scatter in the luminosity measurements around the 
regression line is adopted as a representative uncertainty in 
using single-epoch luminosity measurements to measure the AGN mean 
luminosity (thus justifying the choice in \S~\ref{data}). A higher
uncertainty may apply in practise, especially if (spectro-)photometric 
data are not used to determine the continuum luminosity.

For completeness it is noted that the luminosity offset is not caused by
either (1) interpolation errors in the Neugebauer \et data [tight
spectral energy distributions; interpolation across neighboring pixels],
(2) intrinsic source variability [possibly some contribution, but 
symmetric scatter around a one-to-one relationship would be expected], 
(3) aperture differences between the Kaspi \et and Neugebauer \et studies,
or (4) different corrections for reddening or Galactic extinction.  
As for (3), both studies have apertures including equal amounts of host 
galaxy contributions, which afterall is only strong in a few objects 
(S.\ Kaspi 2001, private communication).  However, the use of slightly 
different absolute flux calibration scales could be the origin of the 
systematic luminosity offset.

\placefigure{LumLum.fig}

\section{Calibration of Single-Epoch Mass Estimates: Optical Measurements 
\label{optcal}}

Having established that the single-epoch line widths and luminosities are 
statistically representative of the multi-epoch equivalent measurements, 
the ``single-epoch central mass estimates'' can be computed and further 
analyzed.
The first step in the calibration of UV single-epoch measurements to estimate
the central AGN/quasar mass is to compare and calibrate, if necessary, the
single-epoch optical measurements to yield representative estimates of the 
central masses.  This is done here using the sample of 18 objects with 
central mass determinations from reverberation mapping (Wandel \et 1999; 
Kaspi \et 2000) and \hb{} measurements. 

\placefigure{M-Mrms.fig}

Using the line widths and luminosities as discussed above the single-epoch 
mass estimates, \mhb, were computed for the 18 AGNs and quasars using 
equations~(\ref{mrl.eq}) and (\ref{blrsize.eq}). These 
masses are plotted in Figure~\ref{M-Mrms.fig} versus the masses determined 
from reverberation mapping studies.  The masses quoted by Kaspi \et (2000) are 
based on the average of the \hb{} and \ha{} line widths. 
It is more appropriate to compare the single-epoch mass estimates with
reverberation masses determined from the \hb{} line width only. 
The reasons are the following:
(1) Kaspi \et determined the BLR size, R$_{\rm BLR}$, using data on both 
\ha{} and \hb{} but found the same results using \hb{} alone, just with 
larger scatter,  (2) The \ha \,$\lambda$\,6563 profile is potentially 
contaminated by [N\,{\sc ii}]\,$\lambda \lambda$\,6548, 6583 which will 
affect the FWHM(\ha) measurement, and, quite importantly, (3) the 
single-epoch mass estimates are based on \hb{} line widths only.
Therefore, the reverberation masses were recomputed using 
equation~(\ref{mrl.eq}) based on FWHM(\hb), measured in the mean and rms
spectra, respectively, and the directly measured BLR sizes, R$_{\rm BLR}$, 
as listed in Tables~6 and 7 by Kaspi \et (2000).
These \hb{} based reverberation masses are listed in Table~\ref{Opt_pars.tab}
along with the single-epoch mass estimates.

\mhbrms{} is plotted in Figure~\ref{M-Mrms.fig}a with the single-epoch mass 
estimates for the 18 objects in sample B (see Table~\ref{regression};
PG\,1351$+$064 has no \hb{} reverberation data). 
Similarly, \mhbmean{} is plotted in Figure~\ref{M-Mrms.fig}b.
The dashed line indicates a one to one relationship with no constant offset 
between the two mass measures.
Some scatter is expected and observed and apart from the usual objects 
with deviating FWHM(\hb) measurements (the outlying objects are labeled),
most of the scatter is likely caused by the intrinsic scatter in the
BLR size $-$ luminosity relationship (see Figure~6 by Kaspi \et 2000). 
This scatter is the cause of PG2130$+$099 being an outlier in these 
diagrams. Disregarding PG1613$+$658, which all along has not behaved 
well, and for a moment the two objects with rather larger uncertainties 
in their mass estimates, PG1700$+$518 and PG1704$+$608, the scatter in 
the \mhbrms $-$ \mbh (\hb, single-epoch) relationship is afterall comparable 
to the uncertainties in the mass estimates.  The best fit regression 
slopes are also plotted in Figure~\ref{M-Mrms.fig}. 
The solid lines are the BCES bisectors when omitting the encircled 
object (PG1613$+$658). The dotted line is the
BCES bisector to all the data points in each diagram.

The results of the linear regressions and the detailed fitting parameters
are listed in Table~\ref{optMregression}. 
The issue of establishing the intrinsic relationship between these two mass 
estimates (single-epoch versus multi-epoch) is very similar to that of the 
most accurate representation of the intrinsic relationship between, say, the 
black hole mass, \mbh, and the stellar velocity dispersion in the host galaxy 
(\eg Gebhardt \et 2000a; Ferrarese \& Merritt 2000; Merritt \& Ferrarese 2001a).  
In both cases the relationship is sought between two 
variables with both measurement uncertainties and intrinsic scatter.
Therefore, for the calibration of the single-epoch mass estimates, the
use of the BCES linear regression algorithm is very important.

The mass comparison of most importance for the calibration 
is that with \mhbrms{} (see \S~\ref{method}).  Since both the single-epoch 
mass estimate and the reverberation mass are (different) measures of the 
same property, the BCES bisector is the most appropriate
regression to use. It is apparent from Table~\ref{optMregression}, however, 
that using either of the BCES (Y$|$X) and bisector regressions yields the 
same basic results within the uncertainties.
Namely, there is a pure unity relationship between 
\mbh (\hb, single-epoch), hereafter \mhbse{}, and \mhbrms{}
within $\lsim$0.25$\sigma$, for BCES(Y$|$X):

\begin{eqnarray}
\log \,{M}_{\rm BH} {\rm (H_{\beta}, rms)} & =&  (0.97\pm0.13) \times
   \log \,M_{\rm BH} (\rm H_{\beta}, S-E) \nonumber \\
    & &     + (0.05 \pm 1.08)
\end{eqnarray}
\noindent
and for the BCES bisector:
\vskip 0.3cm \noindent
\begin{eqnarray}
\log \,{M}_{\rm BH} {\rm (H_{\beta}, rms)} & = & (1.03\pm0.14) \times
   \log \,M_{\rm BH} (\rm H_{\beta}, S-E) \nonumber \\
      & &   - (0.49 \pm 1.17)
\end{eqnarray}
\vskip 0.3cm \noindent
The uncertainties used in the regression are, again, the symmetric 
errors on the logarithms based on the positive linear errors [i.e.,
= $(\log{e})\,(\sigma_X/X)$; \S~\ref{secal}]. 

For completeness the single-epoch mass estimates are also compared 
with the reverberation masses based on the FWHM(\hb, mean) measured 
in the multi-epoch spectra (Fig.~\ref{M-Mrms.fig}b).  The outliers 
(labeled) are again the `usual suspects' (\S~\ref{FWdeviation} and 
above).  Again, there is a unity relationship, but with less scatter 
(Table~\ref{optMregression}).
This is expected as the single-epoch \mbox{FWHM(\hb)} also show less scatter 
with FWHM(\hb, mean) than with FWHM(\hb, rms) [\S~\ref{FWcompar}].  
When comparing the 
single-epoch masses to the masses computed by Kaspi \et (\ie based 
on both \ha{} and \hb) larger scatter and poorer fits are obtained. 
This is expected, as explained, and stresses the need to use the 
reverberation masses based on \hb{} measurements only.

In conclusion, 
the calibration of single-epoch ``optical'' mass estimates to the
multi-epoch reverberation masses is, with the current 
uncertainties, a one-to-one relationship with no significant zero-point 
offset and no apparent reasons to further correct \mhbse{}.
Note, however, that this assumes the line widths are measured as described
in \S~\ref{FWcompar}. 
The performance of the optical single-epoch mass calibration and how it
compares to the  UV single-epoch mass calibration, derived below, are
discussed in \S~\ref{errors}.

\section{Calibration of Single-Epoch Mass Estimates: UV Measurements 
\label{uvcal}}

In this section a calibration is determined of UV spectral measurements 
to reflect reasonable estimates of the central masses.  Two subsets of 
data are used here. The primary sample (UVrev) is the collection of 26 
AGNs and quasars with central mass determinations from reverberation 
mapping (Wandel \et 1999; Kaspi \et 2000) 
for which \civ{} line widths, FWHM(\civ), and UV continuum luminosities, 
$\lambda L_{\lambda}$(1350\AA), are available in the literature. For 
this sample the UV measurements will be directly compared with the 
available reverberation mass determinations, \mhbrms.
By using this sample the UV calibration is based on fewer assumptions 
than if additional objects were included. This is because the 
single-epoch UV mass estimates are compared to independently measured 
central masses and not to masses {\it estimated} based on, for example, 
optical measurements.

Nevertheless, it is instructive to compare the results of the UVrev sample 
with those of a larger sample (named ``sample UV'' in the following). The 
latter sample consists of sample UVrev and 30 other PG quasars for which 
FWHM(\civ) and UV continuum luminosities are likewise readily available 
(Table~\ref{UV_pars.tab}).  For these additional quasars, no reverberation 
masses are available but the central masses are here determined from the 
calibration of single-epoch optical measurements, derived in 
\S~\ref{optcal}. 
It is clear that since these masses are estimates they are expected to 
introduce some uncertainty, which is why they only serve to provide a 
comparison and an additional check on the performance of the calibrations.
However, as discussed later, this uncertainty appears to 
be within the scatter intrinsic to the reverberation masses, which 
indicates that the calibrations are relatively reliable.

In \S~\ref{FWcompar} it was necessary to remeasure the single-epoch 
FWHM(\hb) on the original, uncorrected data (from BG92) to ensure that the 
subtraction of the narrow emission component by BG92 would not affect the 
line width measurement adversely.  This was important for the use of FWHM(\hb) 
to estimate the central mass, as discussed there. Similarly, the single-epoch 
FWHM(\hb)'s were remeasured for the subset of 30 PG quasars, studied by BG92, 
which are used here to extend the primary sample of reverberation mapped AGNs
(UVrev).
And similar to the approach adopted in \S~\ref{FWcompar} each remeasured
single-epoch FWHM(\hb) was compared to the BG92 measurement and corrected
based on the same considerations\footnote{Namely, only the measurements which 
deviated in excess of the assumed 10\% measurement error were corrected.  
For quasars with strong and/or broad 
\feii{} emission but with no significantly strong, spiky narrow component, 
the BG92 measurement was considered the most accurate and fair representation 
of the FWHM(\hb) in the rms spectrum. Quasars which have `uncorrected' 
FWHM(\hb) smaller than BG92 also have relatively strong narrow, sometimes 
spiky, components. In those cases the `uncorrected' FWHM(\hb) was considered 
the most reasonable representation.}.  
Similar to the UVrev objects the optical single-epoch continuum luminosities,
$\lambda L_{\lambda}$(5100\AA), were determined from the tabulated spectral
energy distributions of Neugebauer \et (1987) or the photometry of Schmidt \&
Green (1983). The `optical mass estimates' were then determined from 
equation~(\ref{mrl.eq}) and~(\ref{blrsize.eq}), as justified in \S~\ref{optcal}. 
The adopted optical measurements, FWHM(\hb) and $\lambda 
L_{\lambda}$(5100\AA), and the single-epoch mass estimates, \mhb, are 
listed in Table~\ref{Opt_pars.tab}.

The calibration of the UV measurements to reflect the central mass is described
in the following. First, the mass is expected to depend on FWHM$^2$(\civ).
Anything else would not be physical (according to the virial theorem). Second, one can
assume that the continuum luminosity enters to the power 0.7 as was established
for the optical measurements (Kaspi \et 2000; see also Appendix~\ref{sizelum}
for a discussion thereof), since the continuum luminosities at 5100\,\AA{}
and 1350\,\AA{} are generally expected to be related (via the assumed 
power-law relationship, $L_{\nu} \propto \nu^{-\alpha}$). 
The validity of this latter assumption can be tested by computing the BCES 
regressions between the already established masses and what can be called 
the normalized UV mass, \nMuv = 
FWHM$^2$(\civ) [$\lambda L_{\lambda}$(1350\AA)]$^{0.7}$. The question is
thus whether \nMuv{} is related to \mbh{} to the first power and what is the 
global scaling factor? 
The regression results are presented in Table~\ref{UVMregression} for the two
basic samples, namely sample UVrev and sample UV, and are discussed below. 
In this case where the relationship between \nMuv{} and the reverberation mass,
\mhbrms, will be used 
to calibrate other data it is very important to choose the most appropriate 
statistical method to characterize the intrinsic relationship between these 
two variables.
Different regressions [\eg (Y$|$X), bisector, and orthogonal regressions]
yields statistically different results as they measure different moments of
the data (\eg Isobe \et 1990; Feigelson \& Babu 1992). Which is then the most 
appropriate BCES regression (moment) to use for the calibration? 
This is not clear even from a statistical point of view. 
Therefore, the following approach is adopted. First, BCES regression analyses 
are performed to test whether the relationship between $\log$\,\nMuv{} and 
$\log$\,\mhbrms{} are indeed consistent with a slope of 1.0 to within the 
uncertainties. This justifies the assumption that \mbhuv{} = constant 
$\times$ FWHM$^2$(\civ) [$\lambda L_{\lambda}$(1350\AA)]$^{0.7}$
as discussed above and also justifies the next step. Once this unity
slope is established the problem reduces to the simple relationship with
only one degree of freedom: $ y~ =~ x ~+~ a$ or $a ~=~ y ~-~ x$, where 
both $y$ (= log\,\mhbrms) and $x$ (= log\,\nMuv) have uncertainties. 
The constant zero point offset, $a$, can therefore be determined as the 
weighted mean of ~`$~y ~-~ x~$'~ determined for the individual objects. 
This approach limits the introduction of unnecessary systematic errors
that  would occur if both the slope and intercept are free parameters.
Once $a$ is determined, it can be tested that the calibrated UV estimates, 
\mbhuv = \nMuv $\times 10^a$, are indeed related to the established central 
masses, \mhbrms, in a pure one-to-one relationship (\ie a BCES bisector slope 
of unity and no zero-point offset). 

\placefigure{Muv-Mrms.fig}

Figure~\ref{Muv-Mrms.fig} displays the distribution of \nMuv{} and 
\mhbrms{} for the UVrev sample (panel a) and for the full UV sample
(panel b). The dotted line is the BCES bisector to all the displayed
data points, while the dashed line represents the bisector when NGC\,4151
is excluded (see below). The solid lines are the BCES (Y$|$X) and (X$|$Y) 
regressions to the latter data.
For the UVrev sample alone there is a larger difference between the 
bisector and BCES(Y$|$X) slopes (Fig.~\ref{Muv-Mrms.fig}a; 
Table~\ref{UVMregression}) than for the UV sample 
(Fig.~\ref{Muv-Mrms.fig}b), indicating
the relatively larger scatter in the UVrev sample.  [This is expected, 
however, as the 30 additional PG quasars have smaller intrinsic scatter 
in their optical mass estimates, \mbh(\hb), owing to their origin in the 
mass relationship in eqns.~(\ref{mrl.eq}) and (\ref{blrsize.eq})].
As is also clear from these figures, NGC\,4151 has large uncertainties and 
is very much an outlier. Excluding this data point clearly has a significant 
effect on the slope in both cases (samples UVrev,b and UVb in 
Table~\ref{UVMregression}). The BCES bisector for sample UVrev,b then
shows a slope of 1.0 to within 0.3$\sigma$. The BCES (Y$|$X) regression is 
consistent with unity to within $< 2.5 \sigma$; the uncertainty is somewhat 
large due to the intrinsic scatter in this sample. Note that, excluding also 
PG1704$+$608 and/or PG1613$+$658 does not significantly change the regressions.
When the 30 additional PG quasars are included the statistical significance 
increases and also provides a larger mass range. 
In effect this allows the relationship to be better constrained 
(\eg Fig.~\ref{Muv-Mrms.fig}b). As seen in Table~\ref{UVMregression} for 
sample UV the (Y$|$X) slope is now increased to 0.7 and is improving further 
when NGC\,4151 is excluded (sample UVb). In this case both the BCES (Y$|$X) 
and bisector slopes are consistent with a unity relationship to within at 
most $\sim2 \sigma$, which is acceptable. 
That is, both the UVrev,b and UVb samples show similar relationships in the 
mass comparisons, although with different scatter and uncertainties.

The fact that NGC\,4151 significantly changes the regression results argues 
that it should not be included in the calibration for the same
reason that PG1613$+$658 was excluded from the optical calibration: it likely
behaves differently than the bulk of the AGNs and quasars, and it is thus
inappropriate to apply a calibration to large samples of AGNs that are based
on and significantly affected by (possibly) strangely behaving objects.

Since indeed $\log$\,\nMuv{} and $\log$\,\mhbrms{} are related with slope 
$\equiv$\,1.0, the 
intercept can be determined as the weighted mean of the individual mass 
differences; their uncertainties are propagated from the individual errors 
on \nMuv{} and \mhbrms{}.  For sample UVrev,b the weighted mean is

\begin{equation}
< \log \,{M_{\rm BH}} {\rm (H_{\beta}, rms)} - \log \, nM_{\rm BH, UV} > ~=
	~6.2 \pm 0.03
\end{equation}
\\ 
where the uncertainty is the precision based on the propagated errors.
The sample standard deviation is 0.45 showing the presence of real 
intrinsic scatter, as expected (\eg Fig.~\ref{Muv-Mrms.fig}). 
Excluding the other outliers (PG1613$+$658 and PG1704$+$608) does not 
significantly change the results. 
The weighted mean computed for the full UVb sample is the same, showing
a difference only on the second significant digit (likewise for the 
precision and the sample standard deviation).
The fact that UVrev,b and UVb show very similar results confirms again
that the optical single-epoch mass estimate is a fair representation of
the reverberation masses (based on the current data).

\placefigure{MuvCal-Mhbrms.fig}

How do the calibrated UV estimates, \mbhuv = \nMuv $\times 10^{6.2}$, 
relate to the established central masses, \mhbrms?  
In Figure~\ref{MuvCal-Mhbrms.fig} \mbhuv{} are plotted versus \mhbrms{} 
for the UVrev sample (panel a) and versus \mhbrms{} or \mhbse{} for 
the full UV sample (panel b). The regression results (excluding 
NGC\,4151; Table~\ref{UVMregression}) are also shown. The BCES (Y$|$X), 
(X$|$Y) (dotted lines) and bisector (solid line) regressions, and a 
pure 1:1 relationship (dashed line) are also shown.
The most important comparison here is that with the independently
established reverberation masses in Figure~\ref{MuvCal-Mhbrms.fig}a.
The UV sample is shown in Figure~\ref{MuvCal-Mhbrms.fig}b for completeness.
The BCES bisector fit to the objects in the UVrev,b sample (excluding 
NGC\,4151) yields the following result:

\begin{eqnarray} 
\lefteqn{\log \,M_{\rm BH} {\rm (H_{\beta}, rms)} = } \nonumber \\    
   & & (1.07\pm0.14) \times
   \log \,M_{\rm BH, UV} - 0.61 \pm 1.11 
\label{logMuvcal1.eq}
\end{eqnarray} 

\noindent
The bisector is consistent with a unity relationship to within 
0.5$\sigma$, thereby indicating that the UV calibration is robust.

When the additional 30 PG quasars are included 
(Fig.~\ref{MuvCal-Mhbrms.fig}b) with single-epoch optical mass estimates 
a slightly steeper bisector is found. However, this sample is also 
consistent with a unity relationship in this calibration. 
In this case, it is so to within $\sim 2 \sigma$ and is still
acceptable. Although the two samples have different intrinsic scatter
it is comforting to see that they yield consistent results, as
pointed out earlier.

In conclusion, the best calibration of the UV measurements appear to be

\begin{eqnarray}
\lefteqn{\log \,M_{\rm BH} {\rm (H_{\beta}, rms~or~SE)} = } \nonumber \\
   & & \log \,\left[ \left({\rm \frac{FWHM(C\,IV)}{1000~km~s^{-1}}} \right)^2 ~ 
   \left( \frac{\lambda L_{\lambda} \rm (1350\AA)}{10^{44} \rm ~ergs~s^{-
1}}\right)^{0.7} 
	\right]	\nonumber \\
 & & \mbox{} + 6.2 \pm 0.03 ~~(\pm 0.45)
\label{logMuv.eq}
\end{eqnarray}
\\
The last parenthesis contains the sample standard deviation of the 
weighted mean, which shows the intrinsic scatter in the sample.
As opposed to the uncertainty or precision of 0.03 on this weighted 
mean the standard deviation is probably more representative of the 
uncertainty in the offset for this (UVb) sample. It is also specific 
for the result obtained in the logarithmic representation and makes
little sense when linearized.  Therefore, no uncertainties are 
listed for the linear scaling factor below.
In linear representation, the calibration is:

\begin{eqnarray}
\lefteqn{M_{\rm BH, UV} = } \nonumber \\ 
   & & 1.6\,\times 10^6 
   \left(\rm \frac{FWHM(C\,IV)}{1000~km~s^{-1}} \right)^2 \, 
   \left( \frac{\lambda L_{\lambda} \rm (1350\AA)}{10^{44} \rm ~ergs~s^{-
1}}\right)^{0.7} 
\label{Muv.eq}
\end{eqnarray}

\bigskip
The mass calibrations rely strongly on the size $-$ luminosity 
relationship and its intrinsic scatter dominate the uncertainties
in the mass estimates. Such scatter is naturally expected given the 
nature of the objects and the nature of the obtained BLR sizes based 
on continuum variability (\eg Netzer \& Peterson 1997).
Once a larger sample of AGNs with reverberation mapping masses is
available, the $R_{\rm BLR} - L$ relationship should be updated.
Because it matters how this relationship is established, a slightly
modified approach to that adopted by Kaspi \et (2000) is advocated
in Appendix~\ref{sizelum}. Note that, the modified approach does
not significantly affect the current $R_{\rm BLR} - L$ relationship.

\section{How Reliable Are the Calibrations? \label{errors}} 

It is of keen interest to ask how much of an error we typically will make 
on the mass estimate using the virial mass-luminosity-FWHM relationship 
[eqn~(\ref{mrl.eq}) and (\ref{blrsize.eq})] when instead of the 
FWHM(\hb,rms) and the mean monochromatic luminosity, $\lambda L_{\lambda}$, 
we use FWHM and $\lambda L_{\lambda}$ measurements of a single-epoch 
spectrum, which is in fact a `snap-shot' spectrum at any random given time?  
Clearly, this is only an approximation with the caveat that the estimated 
central mass {\it may} be off by a large factor.  But what is the 
probability for that? In the following, the approximate uncertainties are 
briefly evaluated, assuming that the sample of 26 (18) nearby AGNs and 
quasars studied using reverberation mapping techniques is representative
for the UV (optical) calibration uncertainties.

\placefigure{MCal-problty.fig}

In Figure~\ref{MCal-problty.fig} the reverberation masses, \mhbrms, are 
plotted against the deviation in the optical (panel a) and UV (panel b) 
mass estimates from this established mass.  Offsets of 0.5\,dex and 
1.0\,dex from a perfect one-to-one relationship are indicated. 
The probabilities of estimating the mass with a certain accuracy are
summarized below and in Table~\ref{probabilities.tab}. 
For the optical single-epoch mass estimates Figure~\ref{MCal-problty.fig}a 
shows 17 objects out of the 18 available to have single-epoch mass 
estimates deviating by $\lsim$1.0 dex.  That is, there is a probability 
of $\sim$95\% of getting the mass accurate to within an order of magnitude 
using single-epoch optical spectrophotometry.  Similarly, there is a 
$\sim$90\% probability of getting the mass accurate to within a factor of 6, 
and as much as a $\sim$80\% probability of \mhbse{} being ``correct'' to 
within a factor of 3 (Table~\ref{probabilities.tab}). In other words, the 
1\,$\sigma$ uncertainty in \mhbse{} is a factor of $\lsim$2.5.

Three of the 26 objects in Figure~\ref{MCal-problty.fig}b deviate in 
\mbhuv(\civ, S-E) by more than 1 dex from the reverberation mass, \mhbrms. 
There is, thus, a $\sim$90\% probability of obtaining a central mass, 
accurate to within an order of magnitude, using single-epoch UV spectral 
measurements.  As listed in Table~\ref{probabilities.tab} and illustrated 
in Figure~\ref{MCal-problty.fig}b, the probability of obtaining a mass to 
within a factor of 3 is as high as 70\% (1\,$\sigma$ error), and there is
an even higher chance (85\%) of the mass being within a factor of 6 based 
on the UV calibration presented here.  

The estimated uncertainty intrinsic to the reverberation technique is less 
than a factor\footnote{
The most well established reverberation mass is that of NGC\,5548, which is
based on many broad emission lines.  Peterson \& Wandel (2000) quote an 
uncertainty of $\simeq$0.15\,dex. In comparison, the propagated measurement
errors on \mhbrms{} (Table~\ref{Opt_pars.tab}) range from 0.04\,dex to 1\,dex
(excluding those measurements with $\gsim$100\% errors) with an average 
propagated measurement error of 0.15\,dex. The absolute uncertainty in the 
reverberation masses are difficult to access. If the measurement uncertainties 
of the stellar velocity dispersions are assumed insignificant, the scatter 
in the M $-$ $\sigma$ relationship may indicate the typical uncertainty in the 
reverberation masses. There appears to be a general scatter of $\sim$0.15\,dex 
(Ferrarese \& Merritt 2000), but it can be as much as 0.3\,dex (Ferrarese \et 
2001).  It is clear from Figure~2 by Ferrarese \et that reverberation masses are
no more uncertain than the best masses derived from stellar kinematics.} of 2.
Therefore, it is expected that the mass estimates based on the current optical 
and UV calibrations are not all reliable within as little as a factor of 3.  
Yet it is reassuring that this relatively high accuracy can be achieved in a 
high 70\% $-$ 80\% of the objects (UV and optical calibrations, respectively).

\placefigure{Mdevia-FW.fig}

Figure~\ref{MCal-problty.fig} shows a tendency of the lower-mass objects 
to be overestimated in the single-epoch masses for both the optical and UV 
calibrations.  This may be an ``artifact'' of the objects with very narrow 
and spiky line profiles.  This is demonstrated in Figure~\ref{Mdevia-FW.fig} 
where the same mass deviations are plotted against FWHM(\hb, rms) -- the 
width of the variable \hb{} profile (left panel) --- and against FWHM(\civ, S-E) 
-- the single-epoch \civ{} widths (right panel) ---  since a \civ{} rms width is 
only known for very few of the objects. 
Clearly, the most narrow-lined objects or those varying strongly in their 
narrow line core tend to be overestimated in their central masses. 
This is most pronounced  for the optical mass estimates; note, it is inherit
in the calibration process that the UV masses will scatter almost 
symmetrically about \mhbrms.

The good fortune is that the UV calibration appears to perform better 
for the higher mass and higher luminosity quasars, \ie for FWHM(\civ) 
$>$ 4000\,\kms{} and/or \mbh{} $\gsim 5\times 10^7$\Msol. This is very 
comforting as its primary application will be to samples of high-redshift 
quasars, which are known to be of higher luminosity and are expected to 
occupy the higher end of the black-hole mass range observed 
($\sim 4 \times 10^7 - 1 \times 10^9$\Msol; see also e.g.\  Laor 2000; Ho 
2001).  If the sample of PG quasars are representative most if not all 
quasars can thus have their central masses estimated to within a factor of 
$\sim$5 or less (Figures~\ref{MCal-problty.fig} and~\ref{Mdevia-FW.fig}). 

Since the mass estimates are very sensitive to the line width it is 
important that the signal-to-noise in the measured spectra is sufficiently 
high that the FWHM is measured relatively accurately (e.g., with a
reliable uncertainty of 10\% or less).
This is worth keeping in mind, especially when studying distant, faint quasars.
It still applies, however, that there is a 85\% (90\%) probability that 
the central masses can be estimated to within a factor of 6 based
on UV (optical) data. This is reliable enough for many statistical 
applications.

In conclusion, both the optical and UV single-epoch mass estimates 
indicate approximately similar probabilities of obtaining reliable 
mass estimates with the optical calibration performing slightly better 
in general.  The 1\,$\sigma$ uncertainty is a factor of 3 or better.
The UV calibration performs very well for the higher mass, broader lined 
objects where practically all the mass estimates are within a factor of $\sim$5 
or less of the reverberation mapping masses. 

\section{Summary and Conclusions}

Virial estimates of central black hole masses based on either optical or UV 
single-epoch spectrophotometry are calibrated to recently measured masses of 
nearby AGNs and quasars using reverberation mapping techniques.  
Single-epoch spectral measurements allow the central masses of distant AGNs 
and quasars to be easily estimated. This is important when more direct mass 
measurement techniques cannot easily be applied; moreover, this method 
allows estimates of masses of large samples of AGNs in a short time span.
The following conclusions are reached:

\begin{itemize}
\item
The current data show that the above-mentioned mass estimates are best made as follows:
  \begin{itemize}
      \item
	The signal-to-noise in the single-epoch spectra is high enough that the 
	FWHM can be measured reliably to an accuracy of 10\% or better.
      \item
	The FWHM is measured on a \hb{} or \civ{} emission line
	profile corrected for strong, contaminating \feii{} emission.
	But, for higher luminosity sources (i.e., quasars), narrow-line 
	subtraction should not be attempted.
      \item
	The continuum luminosity 
	measured at 5100\AA{} or
	1350\AA{} is used to estimate the relevant distance of the
	\hb{} or \civ{} emission line gas, respectively, using the
	empirical $R_{\rm BLR} - L$ relationship (Kaspi \et 2000).
  \end{itemize}

\item
The 1\,$\sigma$ uncertainty in the optical and UV calibrations
are a factor of 2.5 and 3, respectively.
  \begin{itemize}
      \item
	Mass estimates based on single-epoch optical spectra are 
	statistically consistent with the reverberation masses to within the
	uncertainties for the current database. 
       The `optical single-epoch mass estimate' measure the reverberation 
	masses to within factors of 3, 6, and 10 with probabilities of 80\%, 
	90\%, and 95\%, respectively.
      \item
       The `UV single-epoch mass estimate' measure the reverberation masses
       to within factors of 3, 6, and 10 with probabilities of 70\%, 85\%, and
       90\%, respectively.
  \end{itemize}

\item
The most deviating mass estimates are found for the lower mass, more 
narrow-lined and/or lower luminosity AGNs. For quasars with FWHM(\civ) 
$>$4000 \kms{} essentially all masses here are accurate to within a 
factor of 3 to $\sim$5. Thus these calibrations seem to perform very well 
where they are most needed, that is at high redshift for higher 
luminosity, more massive, quasars.

\item
The currently obtainable accuracy and associated probabilities are quite
fair given the intrinsic uncertainty (factor less than 2) in the reverberation
masses.
An increase in the accuracy of the calibrations require a smaller scatter
in the BLR size $-$ luminosity relationship. A decrease in the 
measurement uncertainties of the BLR size from reverberation studies
will help. However, it is equally important that the BLR size be 
determined for as large a sample of AGNs and quasars as possible
(and larger than the current sample of 34 objects) spanning a range
of AGN properties.

\end{itemize}

\acknowledgments

It is a great pleasure to thank Luis Ho and Brad Peterson for
encouragement and numerous inspiring discussions from which this work
has benefitted, and their comments on the manuscript.
The author also acknowledges helpful discussions with Matt Bershady and
Rick Pogge.
The original spectra of the PG quasars were kindly provided by Todd Boroson.
The author greatfully acknowledges financial support from the Columbus Fellowship.
This research has made use of the NASA/IPAC Extragalactic Database (NED)
which is operated by the Jet Propulsion Laboratory, California Institute
of Technology, under contract with the National Aeronautics and Space
Administration.

\newpage

\setcounter{figure}{0}
\figcaption[]{ The distribution of FWHM(\hb) in the mean (panel a) and rms
(panel b) multi-epoch spectra and single-epoch FWHM(\hb) of the PG quasars
presented by Kaspi \et (2000) and Mrk\,110 and Mrk\,335 (Wandel \et
1999).  The dotted line indicates a pure one-to-one relationship.
The objects discussed in the text are labeled. The open squares denote
Seyfert~1s while triangles show measurements for the quasars.
The three solid circles show the measurements of Boroson \& Green (1992) which
are based on spectra with both \feii{} emission and the narrow core component
subtracted. These FWHM measurements are not always representative of the BLR
velocity dispersion needed for this study (see text).
The solid line is the best fit BCES bisector regression line based on all
the objects in the diagram except PG1613$+$658 (sample C;
Table~\ref{regression}).  The BCES (Y$|$X) and (X$|$Y) regressions are not
plotted as they crowd the bisector.
The single-epoch FWHM(\hb) scatter around a one-to-one relationship with
FWHM(\hb, mean) to within 15\% $-$ 20\% variation and around a similar
relationship with FWHM(\hb, rms) to within 20\% $-$ 25\%.
Note, the ordinate range is different in the two diagrams.
\label{FWHM.fig}}

\figcaption[]{(a) The distribution of the mean $\lambda L_{\lambda}$(5100\AA)
multi-epoch measurements (Wandel \et 1999; Kaspi \et 2000) based on monitoring
data with
respect to the single-epoch $\lambda L_{\lambda}$(5100\AA) measurements
of Neugebauer \et (1987) and Schmidt \& Green (1983). The errors in the
mean $\lambda L_{\lambda}$(5100\AA) are the rms around this mean. The errors
in the single-epoch $\lambda L_{\lambda}$(5100\AA) are propagated errors
(see text). The short-dashed line (centrally positioned) denotes a one-to-one
relationship. The dotted and long-dashed lines represent $\pm$30\%
and $\pm$60\% luminosity variations, respectively (see text).
(b) The best fit regression lines are shown for the BCES bisector ({\it solid
line}). The BCES (Y$|$X) and (X$|$Y) regression lines would crowd the
bisector, if plotted.  All these BCES regression fits are consistent with a
slope of 1.0. The single-epoch luminosities are offset by $+$0.138 dex at
$\lambda L_{\lambda}$(5100\AA) $\approx$\,44.8 ergs s$^{-1}$, the mid-range
luminosity, based on the BCES bisector.
\label{LumLum.fig}}

\figcaption[]{The reverberation masses derived from the rms (panel a) and
mean (panel b) spectra plotted versus the single-epoch mass estimates
based on optical spectral measurements.
Triangles denote quasars, while squares denote Seyfert\,1s. The dashed
line indicates a unity relationship. The dotted line is a BCES bisector
regression line to all the PG quasars with \hb{} measurements (shown),
while the solid line is the BCES bisector when PG1613 is excluded (see text).
As expected, \mhbse{} show less scatter with \mhbmean. Both relationships
are consistent with a one-to-one relationship to within the errors.
\label{M-Mrms.fig}}

\figcaption[]{ Established and estimated central masses based on optical data
plotted versus the UV measurements. This distribution is the basis of the
calibration of the UV measurements. The ``optical masses'' (ordinate)
in panel (a) consist of the reverberation masses, \mhbrms, ({\it filled, 
circled squares}) derived from the rms spectrum and based on \hb{} only. 
In panel (b) these masses are supplemented with the single-epoch mass 
estimates, \mhbse, ({\it filled squares}) for the 30 PG quasars
with {\it no} reverberation mapping masses.
{\it Dotted lines:} BCES bisector regression lines to all the objects in each diagram.
{\it Solid lines:} BCES(Y$|$X) and (X$|$Y) regressions to all the objects except
NGC\,4151 (see text).
{\it Dashed lines:} the BCES bisector for all objects excluding NGC\,4151.
\label{Muv-Mrms.fig}}

\figcaption[]{
The central mass estimates based on the calibration of the UV spectral
measurements are compared to central masses measured (and/or estimated;
panel b) based on optical data.  {\it Dashed line:} pure unity relationship.
{\it Dotted lines:} BCES(Y$|$X) and BCES(X$|$Y) regression lines. {\it Solid line:}
the BCES bisector. NGC\,4151 was excluded from the regression analysis.
(a) \mbhuv{} estimates versus the established central masses, \mhbrms{}
[UVrev sample only].
(b) \mbhuv{} estimates are plotted for the full UV sample versus the
``optical masses'' described in Figure~\ref{Muv-Mrms.fig}b.
For both diagrams the mass relationships are consistent with
a unity relationship within the uncertainties.
\label{MuvCal-Mhbrms.fig}}

\figcaption[]{The established central masses, \mhbrms, based on optical multi-epoch
spectral measurements plotted versus the deviations of mass estimates based on
calibrated single-epoch spectra. (a) Deviations of masses
estimated from optical spectra, i.e., \mhbse\  divided by \mhbrms. 
(b) Deviations in the masses based on UV spectral measurements, \mbhuv(\civ).
The uncertainties in the abscissa are the (propagated) uncertainties in
the single-epoch masses (\ie {\it not} the mass deviation error).
A strictly unity relationship is indicated by the solid line. Offsets of
$\pm$0.5\,dex ($\pm$1\,dex) are indicated by the dotted (dashed) lines.
\label{MCal-problty.fig}}

\figcaption[]{The mass deviations from Figure~\ref{MCal-problty.fig} plotted here
versus the \hb{} and \civ{} line widths. The optical mass deviations are plotted
against FWHM(\hb, rms), the width of the rms profile (i.e., the variable part),
in (a) and versus the single-epoch FWHM(\civ) in (b).
Lines and symbols are as in Figure~\ref{MCal-problty.fig}.
Notice that the larger mass discrepancies tend to occur for the most narrow
lined objects in both cases (or those with the strongest variability occurring
in the narrow line core: PG1613$+$658 and PG1704$+$608).
\label{Mdevia-FW.fig}}


\begin{appendix}
\section{The BLR Size $-$ Luminosity Relationship Revisited \label{sizelum}}

The aim of this section is twofold, namely (1) to demonstrate the importance
of using regression analysis that is appropriate for the nature of the
data set, and (2) to advocate a careful assessment of the object sample on
which important calibrations, such as the $R_{\rm BLR} - L$ relationship, are 
based.  While this approach yields an $R_{\rm BLR} - L$ relationship consistent 
with equation (2) for the current object sample and data base, it is expected
to become important once more data are available, allowing better constraints
to be placed on this relationship.

Kaspi \et (2000) established the empirical relationship 
[equation~(\ref{blrsize.eq})]
between the BLR size as measured by reverberation mapping and the intrinsic 
continuum luminosity of the objects using linear regression techniques that 
take errors in both variables into account. 
However, as the relationship is sought between two variables that have both 
measurement uncertainties and intrinsic scatter, it is important to use the 
bivariate correlated errors and intrinsic scatter (BCES) algorithm (Akritas 
\& Bershady 1996) to establish the best size $-$ luminosity relationship. The 
results of applying this more appropriate linear regression analysis is shown 
in Table~\ref{RLregression}. 
The regression was done between $\log\,R_{\rm BLR}$ and 
$\log$[$\lambda L_{\lambda}$(5100\AA)/10$^{44} \rm ~ergs~s^{-1}$]
to allow a linear regression. Although $R_{\rm BLR}$ is physically 
speaking a 
function of the luminosity and thus the BCES(Y$|$X) is the appropriate
regression to use, the BCES(X$|$Y) and bisector regression results 
are also listed for completeness. 
The more accurate BLR size of 5.9$^{+3.0}_{-2.0}$ for NGC\,4051 (Peterson \et 
2000) is used here instead of that listed by Kaspi \et (2000), but is not
the cause of the differences discussed below. 

When intrinsic scatter is accounted for the luminosity clearly enters with 
a power different from 0.7 (Table~\ref{RLregression}), 
the luminosity power established by Kaspi et al.  
The linear equivalents of the $\log-\log$ BCES regressions 
are for the sample including NGC\,4051:

\begin{equation}
R_{\rm BLR} = (20.9 \pm 4.7) \left[\frac{\lambda 
L_{\lambda}\rm (5100\AA)}{10^{44} \rm ~ergs~s^{-1}}\right]^{(0.38\pm0.14)}
\label{rl06sy1.eq}
\end{equation}
for the Seyfert~1s alone, and for the PG quasars alone: 

\begin{equation}
R_{\rm BLR} = (41.8 \pm 8.3) \left[\frac{\lambda 
L_{\lambda}\rm (5100\AA)}{10^{44} \rm ~ergs~s^{-1}}\right]^{(0.52\pm0.09)}
\label{rl06pg.eq}
\end{equation}
The linear relationship for the combined sample is:

\begin{equation}
R_{\rm BLR} = (32.4 \pm 4.7) \left[\frac{\lambda 
L_{\lambda}\rm (5100\AA)}{10^{44} \rm ~ergs~s^{-1}}\right]^{(0.58\pm0.09)}
\label{rl06.eq}
\end{equation}
The uncertainties used in the regression are the
symmetric errors on the logarithms based on the positive linear errors.
The errors on the proportionality factor
are the linearized errors of the errors on $\log R_{\rm BLR}$, listed for
comparison with eqn.~\ref{blrsize.eq}. 

When the intrinsic scatter in the data is accounted for the luminosity 
dependence of the BLR size weakens and the uncertainties are increased 
[eqn.~(\ref{rl06.eq})]. However, these larger uncertainties are perhaps 
more representative than those returned by a weighted regression alone 
(eqn.~\ref{blrsize.eq}). Note that the slopes of the individual samples 
[eqn.~(\ref{rl06sy1.eq}) and (\ref{rl06pg.eq})] are consistent within 
their errors.

Peterson \et (2000) find that NGC\,4051 is peculiar as it has a larger BLR 
size for its luminosity than the other objects, deviating by almost 3$\sigma$.
Excluding this object they find a luminosity dependence to the power 0.6 
using a variance weighted regression algorithm similar to Kaspi et al., who 
included all objects in their regression.
When NGC\,4051 is excluded the BCES linear regressions change as follows: 

\begin{equation}
R_{\rm BLR} = (23.4 \pm 4.5) \left[\frac{\lambda 
L_{\lambda}\rm (5100\AA)}{10^{44} \rm ~ergs~s^{-1}}\right]^{(0.56\pm0.26)}
\label{modRblrSey.eq}
\end{equation}
for the sample of Seyferts

\begin{equation}
R_{\rm BLR} = (30.2 \pm 5.6) \left[\frac{\lambda 
L_{\lambda}\rm (5100\AA)}{10^{44} \rm ~ergs~s^{-1}}\right]^{(0.66\pm0.09)}
\label{modRblrPGSey.eq}
\end{equation}
and for the combined PG and Seyfert sample.

The value for NGC\,4051 is seen to have a significant effect on the slope. 
Excluding this object makes the Seyfert sample regression consistent 
with that of the combined sample to within the larger uncertainties.  
It is also consistent with the
result of Peterson \et (2000). Both the modified regression slopes, shown 
in their linear representations in equations~(\ref{modRblrSey.eq}) and 
(\ref{modRblrPGSey.eq}), also agree with the result by Kaspi \et (2000) 
within the uncertainties.  
Which slope should be used to estimate the central masses? 
Similar to the mass calibrations (\S\S~\ref{optcal} and \ref{uvcal}) 
one can argue that a calibration like the present should not include an 
extreme outlier that may not be representative of the common AGN,
especially when it has a significant effect on the result.
Therefore equation~(\ref{modRblrPGSey.eq}) seems the most appropriate 
size $-$ luminosity relationship to use. However, given the current scatter 
and uncertainties, using a slope of 0.7 and a scaling factor of 32.9
makes little difference to the mass estimates. However, the larger 
uncertainties 
in equation~(\ref{modRblrPGSey.eq}) are
more representative than those of eqn.~(\ref{blrsize.eq}).

Note, the regression results do not change significantly from those 
listed when the linear negative errors on the BLR size is used instead 
of the positive ones, or if the largest of the two is used for each 
object; this in fact illustrates the robustness of the BCES regression.

The precise value of the luminosity power in the size $-$ luminosity
relationship is important for the single-epoch mass calibrations.
Mass estimates of distant AGNs are highly dependent thereon as the 
masses cannot at present be determined independently in a direct 
manner. This stresses the dire need to obtain BLR size measurements 
for a larger sample of AGNs and to a higher degree of accuracy such 
that this relationship can be further constrained and studied.
Once such data are available the approach presented here (use of
BCES regression analysis and exclusion of objects, such as NGC\,4051,
which are not good representations of the typical AGN population) 
should be adopted.


\placetable{RLregression} 

\setcounter{table}{0}

\begin{deluxetable}{lllccc}  
\tablewidth{0pt}
\hspace{-2cm}
\tablecaption{BCES Regression for BLR Size $-$ Luminosity Relationship \label{RLregression}}
\tablehead{
\colhead{\bf Independent} &
\colhead{\bf Dependent} &
\colhead{\bf Sam-} &
\colhead{\bf N} &
\colhead{\bf Slope} &
\colhead{\bf Intercept} \\
\colhead{\bf Variable} &
\colhead{\bf Variable} &
\colhead{\bf ple} & 
\colhead{(\#)} &
\colhead{\bf $\pm$error} &
\colhead{\bf $\pm$error} 
}

\tablecolumns{6}
\startdata
 \multicolumn{6}{c}{\bf Bivariate Errors and Intrinsic Scatter Regression (BCES) } \\[3pt]
\tableline \\[-9pt]
$\log$R\tablenotemark{a} & $\log \lambda$L$_{\lambda}$\tablenotemark{b} & All & 34 & {\bf 0.58$\pm$0.09 }&{\bf 1.51$\pm$0.06 } \\
$\log \lambda$L$_{\lambda}$ & $\log$R & All & 34 & 0.63$\pm$0.11 & 1.51$\pm$0.06 \\
  \multicolumn{2}{l}{BCES bisector}   & All & 34 & 0.61$\pm$0.09 & 1.51$\pm$0.06 \\
\tableline \tableline \\[-9pt]
$\log$R & $\log \lambda$L$_{\lambda}$ & Sy~1s & 17 & {\bf 0.38$\pm$0.14} & {\bf1.32$\pm$0.10 } \\
$\log \lambda$L$_{\lambda}$ & $\log$R & Sy~1s & 17 & 0.56$\pm$0.25 & 1.41$\pm$0.15  \\
  \multicolumn{2}{l}{BCES bisector}   & Sy~1s & 17 & 0.47$\pm$0.17 & 1.36$\pm$0.11  \\
\tableline \\[-9pt]
$\log$R & $\log \lambda$L$_{\lambda}$ & PGs   & 17 & {\bf 0.52$\pm$0.09} & {\bf1.62$\pm$0.09 } \\
$\log \lambda$L$_{\lambda}$ & $\log$R & PGs   & 17 & 0.45$\pm$0.32 & 1.67$\pm$0.21  \\
  \multicolumn{2}{l}{BCES bisector}   & PGs   & 17 & 0.49$\pm$0.20 & 1.64$\pm$0.18  \\
\tableline \tableline \\[-9pt]
$\log$R\tablenotemark{a} & $\log \lambda$L$_{\lambda}$\tablenotemark{b} & A$^{\prime}$ & 33 & {\bf 0.66$\pm$0.09} & {\bf1.48$\pm$0.07   }    \\
$\log \lambda$L$_{\lambda}$ & $\log$R & A$^{\prime}$ & 33 & 0.71$\pm$0.12 & 1.47$\pm$0.06  \\
  \multicolumn{2}{l}{BCES bisector}   & A$^{\prime}$ & 33 & 0.68$\pm$0.10 & 1.47$\pm$0.06  \\
\tableline \tableline \\[-9pt]
$\log$R & $\log \lambda$L$_{\lambda}$ & S$^{\prime}$ & 16 &{\bf 0.56$\pm$0.26} &{\bf 1.37$\pm$0.10 } \\
$\log \lambda$L$_{\lambda}$ & $\log$R & S$^{\prime}$ & 16 & 0.77$\pm$0.26 & 1.46$\pm$0.15  \\
  \multicolumn{2}{l}{BCES bisector}   & S$^{\prime}$ & 16 & 0.66$\pm$0.19 & 1.41$\pm$0.10  \\
\tableline \tableline \\[-9pt]
\multicolumn{6}{l}{Sample All: All Seyfert~1s and PG quasars (ref [1,2])} \\
\multicolumn{6}{l}{Sample Sy~1s: The Seyfert~1s only (ref [1,2])} \\ 
\multicolumn{6}{l}{Sample PGs: The PG quasars (only) (ref [2]) } \\
\multicolumn{6}{l}{Sample A$^{\prime}$: All Seyfert~1s and PG quasars except NGC\,4051. See text } \\
\multicolumn{6}{l}{Sample S$^{\prime}$: The Seyfert~1s except NGC\,4051. See text}  \\
\tableline 
\enddata
\tablenotetext{a}{$\log$[(BLR Size, R$_{\rm BLR}$)/ light-days]}
\tablenotetext{b}{$\log$[$\lambda$L$_{\lambda}$(5100\,\AA)/ 10$^{44}$ ergs s$^{-1}$] }
\tablerefs{
(1) Wandel \et 1999;
(2) Kaspi \et 2000
}
\end{deluxetable}

\end{appendix}

\clearpage

\setcounter{table}{0}

\vspace{-7cm}
\hspace{-5cm}
\begin{deluxetable}{llccccccccc} 
\vspace{-15cm}
\tablewidth{0pt}
\hspace{-14cm}
\tablecaption{Optical Spectral Parameters and Masses \label{Opt_pars.tab}}
\tabletypesize{\tiny}
\tablehead{
\colhead{\bf Object} &
\colhead{\bf Alt. } &
\colhead{\bf Red-} &
\colhead{\bf FWHM\tablenotemark{a}} &
\colhead{\bf Ref.} &
\colhead{\bf log\,[$\lambda$L$_{\lambda}$]\tablenotemark{b}} &
\colhead{\bf Ref.} &
\colhead{\bf log\,[M/M$_{\bf \odot}$]\tablenotemark{c}} &
\colhead{\bf log\,[M/M$_{\bf \odot}$]\tablenotemark{d}} &
\colhead{\bf log\,[M/M$_{\bf \odot}$]\tablenotemark{d}} &
\colhead{\bf Ref.} \\
\colhead{\bf } &
\colhead{\bf Name} &
\colhead{\bf shift} &
\colhead{\bf (H$\bf \beta$,S-E)} &
\colhead{\bf } &
\colhead{\bf } &
\colhead{\bf } &
\colhead{\bf (H$\bf \beta$,S-E)} &
\colhead{\bf (H$\bf \beta$, rms)} &
\colhead{\bf (H$\bf \beta$, mean)} &
\colhead{\bf } 
}

\tablecolumns{13}
\startdata
%
3C\,120          &      {} & 0.033  & \nodata & \nodata & \nodata & \nodata &   \nodata &     7.48 $^{+0.21}_{-0.27}$ &    7.36 $^{+0.22}_{-0.28}$ & 1 \\ 
3C\,390.3        &      {} & 0.057  & \nodata & \nodata & \nodata & \nodata &   \nodata &     8.57 $^{+0.12}_{-0.21}$ &    8.53 $^{+0.12}_{-0.21}$ & 1 \\ 
Akn\,120         &      {} & 0.033  & \nodata & \nodata & \nodata & \nodata &   \nodata &     8.27 $^{+0.08}_{-0.12}$ &    8.26 $^{+0.08}_{-0.12}$ & 1 \\ 
Fairall\,9       &      {} & 0.046  & \nodata & \nodata & \nodata & \nodata &   \nodata &     7.92 $^{+0.11}_{-0.32}$ &    7.90 $^{+0.11}_{-0.31}$ & 1 \\ 
Mrk\,79          &      {} & 0.022  & \nodata & \nodata & \nodata & \nodata &   \nodata &     8.01 $^{+0.14}_{-0.35}$ &    7.72 $^{+0.14}_{-0.34}$ & 1 \\ 
Mrk\,509         &      {} & 0.035  & \nodata & \nodata & \nodata & \nodata &   \nodata &     7.96 $^{+0.05}_{-0.06}$ &    7.76 $^{+0.05}_{-0.05}$ & 1 \\ 
NGC\,4151        &      {} & 0.003  & \nodata & \nodata & \nodata & \nodata &   \nodata &     7.08 $^{+0.23}_{-0.38}$ &    7.18 $^{+0.23}_{-0.38}$ & 1 \\ 
NGC\,3783        &      {} & 0.010  & \nodata & \nodata & \nodata & \nodata &   \nodata &     7.04 $^{+0.30}_{-0.96}$ &    6.97 $^{+0.30}_{-0.97}$ & 1 \\ 
NGC\,5548        &      {} & 0.017  & \nodata & \nodata & \nodata & \nodata &   \nodata &     7.97 $^{+0.07}_{-0.07}$ &    8.09 $^{+0.07}_{-0.07}$ & 1 \\ 
NGC\,7469   &      {} & 0.017  & \nodata & \nodata & \nodata & \nodata &   \nodata &     6.88 $^{+0.30}_{-6.88}$ &    6.81 $^{+0.30}_{-6.81}$ & 1 \\ 
\tableline
PG\,0003$+$199   & Mrk\,335& 0.025  &   1640 $\pm$ \phantom{0}164 & 2 &   44.00 $\pm$ 0.13 & 5 &     7.14 $\pm$ 0.12 &    6.58 $^{+0.14}_{-0.14}$ &    6.80 $^{+0.12}_{-0.09}$ & 1 \\
PG\,0007$+$106   &III\,Zw\,2& 0.089  &   3979 $\pm$ \phantom{0}398 & 3 &   44.59 $\pm$ 0.07 & 4 &     8.29 $\pm$ 0.10 &   \nodata                 &   \nodata                 &   \\
PG\,0026$+$129   &      {} & 0.142  &   1860 $\pm$ \phantom{0}250 & 2 &   44.90 $\pm$ 0.07 & 4 &     7.85 $\pm$ 0.12 &    7.48 $^{+0.08}_{-0.11}$ &    7.86 $^{+0.08}_{-0.11}$ & 6 \\
PG\,0050$+$124   &I\,Zw\,1 & 0.061  &   1240 $\pm$ \phantom{0}124 & 2 &   44.52 $\pm$ 0.12 & 5 &     7.26 $\pm$ 0.11 &   \nodata                 &   \nodata                 &   \\
PG\,0052$+$251   &      {} & 0.155  &   5200 $\pm$ \phantom{0}520 & 2 &   44.87 $\pm$ 0.07 & 4 &     8.72 $\pm$ 0.10 &    8.61 $^{+0.10}_{-0.10}$ &    8.49 $^{+0.10}_{-0.10}$ & 6 \\
PG\,0157$+$001   &Mrk\,1014& 0.164  &   2460 $\pm$ \phantom{0}320 & 2 &   44.81 $\pm$ 0.07 & 4 &     8.03 $\pm$ 0.11 &   \nodata                 &   \nodata                 &   \\
PG\,0804$+$761   &      {} & 0.100  &   3070 $\pm$ \phantom{0}307 & 2 &   44.93 $\pm$ 0.07 & 4 &     8.31 $\pm$ 0.10 &    8.13 $^{+0.04}_{-0.04}$ &    8.31 $^{+0.04}_{-0.04}$ & 6 \\
PG\,0844$+$349   &TON\,951 & 0.064  &   2420 $\pm$ \phantom{0}242 & 2 &   44.34 $\pm$ 0.07 & 4 &     7.69 $\pm$ 0.10 &    7.45 $^{+0.15}_{-0.21}$ &    7.42 $^{+0.15}_{-0.21}$ & 6 \\
PG\,0921$+$525   &Mrk\,110 & 0.035  &   1816 $\pm$ \phantom{0}182 & 3 &   43.55 $\pm$ 0.12 & 5 &     6.91 $\pm$ 0.12 &    6.89 $^{+0.14}_{-0.21}$ &    6.75 $^{+0.13}_{-0.19}$ & 1 \\
PG\,0923$+$129   &Mrk\,705 & 0.029  &   1742 $\pm$ \phantom{0}248 & 3 &   43.66 $\pm$ 0.13 & 5 &     6.95 $\pm$ 0.14 &   \nodata                 &   \nodata                 &   \\
PG\,0953$+$414   &K348-7 & 0.239  &   3130 $\pm$ \phantom{0}470 & 2 &   45.30 $\pm$ 0.06 & 4 &     8.58 $\pm$ 0.13 &    8.22 $^{+0.06}_{-0.09}$ &    8.27 $^{+0.06}_{-0.09}$ & 6 \\
PG\,1012$+$008   &      {} & 0.185  &   2640 $\pm$ \phantom{0}264 & 2 &   44.83 $\pm$ 0.07 & 4 &     8.11 $\pm$ 0.10 &   \nodata                 &   \nodata                 &   \\
PG\,1049$-$005   &      {} & 0.357  &   5360 $\pm$ \phantom{0}536 & 2 &   45.43 $\pm$ 0.06 & 4 &     9.14 $\pm$ 0.10 &   \nodata                 &   \nodata                 &   \\
PG\,1100$+$772   &3C\,249.1& 0.313  &   4973 $\pm$ \phantom{0}497 & 3 &   45.41 $\pm$ 0.06 & 4 &     9.07 $\pm$ 0.10 &   \nodata                 &   \nodata                 &   \\
PG\,1103$-$006   & PKS     & 0.425  &   6190 $\pm$ \phantom{0}619 & 2 &   45.47 $\pm$ 0.06 & 4 &     9.29 $\pm$ 0.10 &   \nodata                 &   \nodata                 &   \\
PG\,1114$+$445   &      {} & 0.144  &   4570 $\pm$ \phantom{0}457 & 2 &   44.59 $\pm$ 0.07 & 4 &     8.41 $\pm$ 0.09 &   \nodata                 &   \nodata                 &   \\
PG\,1116$+$215   &TON\,1388& 0.177  &   2920 $\pm$ \phantom{0}292 & 2 &   45.27 $\pm$ 0.07 & 4 &     8.50 $\pm$ 0.10 &   \nodata                 &   \nodata                 &   \\
PG\,1119$+$120   &Mrk\,734 & 0.049  &   1820 $\pm$ \phantom{0}182 & 2 &   44.00 $\pm$ 0.08 & 4 &     7.20 $\pm$ 0.10 &   \nodata                 &   \nodata                 &   \\
PG\,1202$+$281   &GQ\,COM & 0.165  &   3715 $\pm$ \phantom{0}372 & 3 &   44.42 $\pm$ 0.07 & 4 &     8.12 $\pm$ 0.09 &   \nodata                 &   \nodata                 &   \\
PG\,1211$+$143   &      {} & 0.085  &   1860 $\pm$ \phantom{0}186 & 2 &   44.94 $\pm$ 0.07 & 4 &     7.88 $\pm$ 0.10 &    7.51 $^{+0.10}_{-0.16}$ &    7.70 $^{+0.09}_{-0.15}$ & 6 \\
PG\,1216$+$069   &      {} & 0.334  &   5190 $\pm$           1020 & 2 &   45.50 $\pm$ 0.06 & 4 &     9.17 $\pm$ 0.15 &   \nodata                 &   \nodata                 &   \\
PG\,1226$+$023   & 3C\,273 & 0.158  &   3520 $\pm$ \phantom{0}352 & 2 &   45.89 $\pm$ 0.07 & 4 &     9.10 $\pm$ 0.11 &    8.63 $^{+0.06}_{-0.06}$ &    8.82 $^{+0.06}_{-0.06}$ & 6 \\
PG\,1229$+$204   &TON\,1542& 0.064  &   3360 $\pm$ \phantom{0}336 & 2 &   44.27 $\pm$ 0.07 & 4 &     7.93 $\pm$ 0.10 &    7.95 $^{+0.17}_{-0.27}$ &    7.94 $^{+0.17}_{-0.27}$ & 6 \\
PG\,1244$+$026   &      {} & 0.048  &    830 $\pm$ \phantom{000}83 & 2 &  43.67 $\pm$ 0.08 & 4 &     6.29 $\pm$ 0.10 &   \nodata                 &   \nodata                 &   \\
PG\,1259$+$593   & LB\,2522 & 0.472  &   3390 $\pm$           1139 & 2 &   45.77 $\pm$ 0.06 & 4 &     8.98 $\pm$ 0.23 &   \nodata                 &   \nodata                 &   \\
PG\,1302$-$102   & PKS     & 0.286  &   3400 $\pm$ \phantom{0}340 & 2 &   45.70 $\pm$ 0.06 & 4 &     8.94 $\pm$ 0.10 &   \nodata                 &   \nodata                 &   \\
PG\,1307$+$085   &      {} & 0.155  &   5320 $\pm$ \phantom{0}532 & 2 &   44.86 $\pm$ 0.07 & 4 &     8.73 $\pm$ 0.10 &    8.70 $^{+0.14}_{-0.46}$ &    8.50 $^{+0.14}_{-0.46}$ & 6 \\
PG\,1309$+$355   &TON\,1565 & 0.184  &   2940 $\pm$ \phantom{0}476 & 2 &   44.84 $\pm$ 0.07 & 4 &     8.21 $\pm$ 0.13 &   \nodata                 &   \nodata                 &   \\
PG\,1351$+$640   &      {} & 0.087  &   5660 $\pm$ \phantom{0}566 & 2 &   44.71 $\pm$ 0.07 & 4 &     8.69 $\pm$ 0.10 &   \nodata                 &   \nodata                 &   \\
PG\,1352$+$183   &PB\,4142 & 0.158  &   3600 $\pm$ \phantom{0}360 & 2 &   44.68 $\pm$ 0.07 & 4 &     8.27 $\pm$ 0.09 &   \nodata                 &   \nodata                 &   \\
PG\,1411$+$442   &PB\,1732 & 0.089  &   2670 $\pm$ \phantom{0}267 & 2 &   44.50 $\pm$ 0.07 & 4 &     7.89 $\pm$ 0.10 &    8.05 $^{+0.14}_{-0.20}$ &    7.96 $^{+0.14}_{-0.20}$ & 6 \\
PG\,1415$+$451   &      {} & 0.114  &   2620 $\pm$ \phantom{0}262 & 2 &   44.42 $\pm$ 0.07 & 4 &     7.81 $\pm$ 0.09 &   \nodata                 &   \nodata                 &   \\
PG\,1416$-$129   &      {} & 0.129  &   3766 $\pm$ \phantom{0}377 & 3 &   44.95 $\pm$ 0.07 & 4 &     8.50 $\pm$ 0.10 &   \nodata                 &   \nodata                 &   \\
PG\,1426$+$015   &Mrk\,1383& 0.086  &   5940 $\pm$ \phantom{0}594 & 3 &   44.74 $\pm$ 0.07 & 4 &     8.75 $\pm$ 0.10 &    8.63 $^{+0.13}_{-0.24}$ &    8.73 $^{+0.13}_{-0.24}$ & 6 \\
PG\,1440$+$356   &Mrk\,478 & 0.077  &   1450 $\pm$ \phantom{0}145 & 2 &   44.42 $\pm$ 0.07 & 4 &     7.30 $\pm$ 0.10 &   \nodata                 &   \nodata                 &   \\
PG\,1444$+$406   &      {} & 0.267  &   2480 $\pm$ \phantom{0}248 & 2 &   45.08 $\pm$ 0.06 & 4 &     8.23 $\pm$ 0.10 &   \nodata                 &   \nodata                 &   \\
PG\,1501$+$106   &Mrk\,841 & 0.036  &   3465 $\pm$ \phantom{0}347 & 3 &   44.17 $\pm$ 0.09 & 4 &     7.88 $\pm$ 0.10 &   \nodata                 &   \nodata                 &   \\
PG\,1512$+$370   &4C\,37.43& 0.371  &   4352 $\pm$ \phantom{0}435 & 3 &   45.41 $\pm$ 0.06 & 4 &     8.95 $\pm$ 0.10 &   \nodata                 &   \nodata                 &   \\
PG\,1534$+$580   & Mrk\,290& 0.030  &   3060 $\pm$ \phantom{0}306 & 3 &   43.58 $\pm$ 0.09 & 4 &     7.36 $\pm$ 0.10 &   \nodata                 &   \nodata                 &   \\
PG\,1545$+$210   &3C\,323.1& 0.266  &   5961 $\pm$ \phantom{0}596 & 3 &   45.26 $\pm$ 0.06 & 4 &     9.12 $\pm$ 0.10 &   \nodata                 &   \nodata                 &   \\
PG\,1613$+$658   &Mrk\,876 & 0.129  &   8450 $\pm$ \phantom{0}845 & 2 &   44.65 $\pm$ 0.07 & 4 &     8.99 $\pm$ 0.10 &    7.55 $^{+0.18}_{-0.20}$ &    8.45 $^{+0.18}_{-0.20}$ & 6 \\
PG\,1704$+$608   & 3C\,351 & 0.371  &   1180 $\pm$ \phantom{0}150 & 3 &   45.50 $\pm$ 0.06 & 4 &     7.88 $\pm$ 0.12 &    6.87 $^{+0.26}_{-1.12}$ &    7.57 $^{+0.27}_{-1.03}$ & 6 \\
PG\,2130$+$099   &II\,Zw\,136& 0.061  &   2330 $\pm$ \phantom{0}233 & 2 &   44.37 $\pm$ 0.07 & 4 &     7.68 $\pm$ 0.10 &    8.42 $^{+0.13}_{-0.07}$ &    8.23 $^{+0.13}_{-0.07}$ & 6 \\
PG\,2209$+$184   &      {} & 0.070  &   6500 $\pm$ \phantom{0}650 & 2 &   44.30 $\pm$ 0.07 & 4 &     8.52 $\pm$ 0.10 &   \nodata                 &   \nodata                 &   \\
PG\,2251$+$113   &  PKS    & 0.323  &   4160 $\pm$ \phantom{0}710 & 2 &   45.48 $\pm$ 0.06 & 4 &     8.96 $\pm$ 0.14 &   \nodata                 &   \nodata                 &   \\
PG\,2308$+$098   & 4C\,09.72 & 0.432  &   7920 $\pm$ \phantom{0}792 & 2 &   45.52 $\pm$ 0.11 & 5 &     9.57 $\pm$ 0.12 &   \nodata                 &   \nodata                 &   \\
\enddata
\tablenotetext{a}{FWHM(\hb) measured in the single-epoch spectrum in units of km s$^{-1}$.}
\tablenotetext{b}{$\log$\,[$\lambda L_{\lambda}$(5100\AA)/ergs s$^{-1}$].}
\tablenotetext{c}{The central mass (and uncertainties) estimated based on single-epoch optical
        spectroscopy.}
\tablenotetext{d}{The central mass (and uncertainties) determined from multi-epoch
        spectrophotometry: FWHM(\hb) in the rms and mean spectrum, respectively,
        and R$_{\rm BLR} = c\tau_{\rm cent}$, corrected to the restframe; these
        measurements are determined using reverberation mapping techniques.
}
\vspace{-0.2cm}
\tablerefs{
(1) Wandel, Peterson, \& Malkan 1999;
(2) Boroson \& Green 1992;
(3) This author's own measurements of the original, {\it uncorrected} spectra presented by ref.[2];
(4) Neugebauer \et 1987;
(5) Schmidt \& Green 1983;
(6) Kaspi \et 2000.
}
\end{deluxetable}

\begin{deluxetable}{lcccll} 
\tablewidth{490pt}
\hspace{-14cm}
\tablecaption{Ultraviolet Spectral Parameters and Masses \label{UV_pars.tab}}
\tabletypesize{\footnotesize}
\tablecolumns{6}
\tablehead{
\colhead{\bf Object\phantom{mmmmm}} &
\colhead{\bf \phantom{mml}Redshift\phantom{mmm}} &
\colhead{\bf \phantom{mm}FWHM\tablenotemark{a}\phantom{mm}} &
\colhead{\bf \phantom{mmm}Ref.\tablenotemark{b}\phantom{mmm}} &
\colhead{\bf log\,[$\lambda$L$_{\lambda}$]\tablenotemark{c}\phantom{mmmm}} &
\colhead{\bf log\,[M/M$_{\bf \odot}$]\tablenotemark{d}} \\
\colhead{\bf } &
\colhead{\bf } &
\colhead{\bf \phantom{mml}(C{\footnotesize IV}, S-E)\phantom{mm}} &
\colhead{\bf } &
\colhead{\bf } &
\colhead{\bf (C{\footnotesize IV}, S-E)} 
}
\startdata
%
3C\,120          & 0.033  &  3807 $\pm$  380 & 1     &   44.17 $\pm$   0.09          &  7.48 $\pm$ 0.12  \\ 
3C\,390.3        & 0.057  &  7700 $\pm$  730 & 2     &   43.74 $\pm$   0.17          &  7.79 $\pm$ 0.17  \\ 
Akn\,120         & 0.033  &  4908 $\pm$  802 & 1,3   &   44.70 $\pm$   0.12          &  8.07 $\pm$ 0.17  \\ 
Fairall\,9       & 0.046  &  3826 $\pm$  805 & 1,3   &   44.88 $^{+~0.09}_{-~0.11}$    &  7.98 $\pm$ 0.20 \\ 
Mrk\,79          & 0.022  &  5552 $\pm$  560 & 1     &   43.87 $\pm$   0.11          &  7.60 $\pm$ 0.13 \\ 
Mrk\,509         & 0.035  &  5303 $\pm$  658 & 1,3   &   44.46 $^{+~0.08}_{-~0.09}$    &  7.97 $\pm$0.13 \\ 
NGC\,4151        & 0.003  &  2190 $\pm$  562 & 3     &   41.36 $^{+~0.29}_{-~0.60}$    &  5.03 $^{+~0.37}_{-~0.32}$ \\ 
NGC\,3783        & 0.010  &  3476 $\pm$  336 & 4     &   43.10 $\pm$   0.15          &  6.65 $\pm$ 0.15  \\ 
NGC\,5548  (mean)& 0.017  &  6422 $\pm$ 2529 & 1,3,5,6 & 43.64 $^{+~0.11}_{-0.~16}$    &  7.56 $\pm$ 0.36 \\ 
$-$\iue~(1989)   & 0.017  &  5520 $\pm$  380 & 5     &   43.51 $\pm$   (0.10)        &  7.34 $\pm$ 0.10  \\ 
$-$\HST~(1993)   & 0.017  &  8950 $\pm$  570 & 6     &   43.43 $\pm$   (0.10)        &  7.70 $\pm$ 0.10  \\ 
NGC\,7469        & 0.017  &  4570 $\pm$  380 & 7     &   43.49 $\pm$   0.08          &  7.16 $\pm$ 0.10  \\ 
\tableline
PG\,0003$+$199   & 0.025  &  3777 $\pm$  375 & 1       &   44.23 $\pm$   0.10        &  7.52 $\pm$ 0.12  \\ 
PG\,0007$+$106   & 0.089  &  4390 $\pm$  730 & F1,8,L1 &   44.72 $\pm$   0.11        &  7.99 $\pm$ 0.17  \\ 
PG\,0026$+$129   & 0.142  &  3261 $\pm$  527 & F1,8,L1 &   45.11 $\pm$   0.07        &  8.00 $\pm$ 0.15  \\ 
PG\,0050$+$124   & 0.061  &  1885 $\pm$  190 & 9       &   44.65 $\pm$   0.05        &  7.21 $\pm$ 0.10  \\ 
PG\,0052$+$251   & 0.155  &  6653 $\pm$  665 & 1       &   45.29 $\pm$   0.12        &  8.75 $\pm$ 0.13  \\ 
PG\,0157$+$001   & 0.164  &  4184 $\pm$  420 & 1       &   45.07 $\pm$   0.09        &  8.19 $\pm$ 0.12  \\ 
PG\,0804$+$761   & 0.100  &  4241 $\pm$  203 & F1,8,L1 &   45.07 $\pm$   0.08        &  8.20 $\pm$ 0.08  \\ 
PG\,0844$+$349   & 0.064  &  4439 $\pm$  149 & F1,8,L1 &   44.57 $\pm$   0.01        &  7.89 $\pm$ 0.05  \\ 
PG\,0921$+$525   & 0.035  &  3782 $\pm$  380 & 1       &   43.71 $\pm$   0.16        &  7.15 $\pm$ 0.17  \\ 
PG\,0923$+$129   & 0.029  &  4103 $\pm$  410 & 1       &   43.74 $\pm$   0.15        &  7.24 $\pm$ 0.15  \\ 
PG\,0953$+$414   & 0.239  &  3067 $\pm$  310 & 10      &   45.51 $\pm$   (0.10)      &  8.23 $\pm$ 0.12  \\ 
PG\,1012$+$008   & 0.185  &  5816 $\pm$  120 & F1,8,L1 &   44.82 $\pm$   0.09        &  8.30 $\pm$ 0.08  \\ 
PG\,1049$-$005   & 0.357  &  3675 $\pm$  450 & F11,L14 &   45.75 $\pm$   0.11        &  8.56 $\pm$ 0.14  \\ 
PG\,1100$+$772   & 0.313  &  8664 $\pm$  900 & F1,8,L1 &   45.60 $\pm$   0.09        &  9.20 $\pm$ 0.12  \\ 
PG\,1103$-$006   & 0.425  &  3500 $\pm$  450 & F11,L14 &   45.87 $\pm$   0.11        &  8.60 $\pm$ 0.15  \\ 
PG\,1114$+$445   & 0.144  &  2246 $\pm$  225 & 1       &   44.70 $\pm$   0.34        &  7.39 $\pm$ 0.37  \\ 
PG\,1116$+$215   & 0.177  &  5329 $\pm$  835 & F8,12,L12 & 45.61 $\pm$   (0.10)      &  8.78 $\pm$ 0.16  \\ 
PG\,1119$+$120   & 0.049  &  5425 $\pm$  545 & 1       &   44.09 $\pm$   0.10        &  7.73 $\pm$ 0.12  \\ 
PG\,1202$+$281   & 0.165  &  2941 $\pm$  290 & F8,12,L12 & 44.27 $\pm$   (0.10)      &  7.33 $\pm$ 0.12  \\ 
PG\,1211$+$143   & 0.085  &  2981 $\pm$  247 & F1,8,L1 &   45.01 $\pm$   0.08        &  7.86 $\pm$ 0.10  \\ 
PG\,1216$+$069   & 0.334  &  3039 $\pm$  300 & 12      &   45.62 $\pm$   (0.10)      &  8.30 $\pm$ 0.12  \\ 
PG\,1226$+$023   & 0.158  &  4141 $\pm$  381 & F8,10,L10 & 46.34 $\pm$   (0.10)      &  9.07 $\pm$ 0.11  \\ 
PG\,1229$+$204   & 0.064  &  4904 $\pm$  490 & 1       &   44.59 $\pm$   0.08        &  7.99 $\pm$ 0.11  \\ 
PG\,1244$+$026   & 0.048  &  2531 $\pm$  255 & 1       &   43.69 $\pm$   0.13        &  6.79 $\pm$ 0.14  \\ 
PG\,1259$+$593   & 0.472  &  5925 $\pm$  800 & F11,L14 &   46.05 $\pm$   0.11        &  9.18 $\pm$ 0.15  \\ 
PG\,1302$-$102   & 0.286  &  3634 $\pm$  365 & 12      &   45.90 $\pm$   (0.10)      &  8.65 $\pm$ 0.12  \\ 
PG\,1307$+$085   & 0.155  &  6399 $\pm$ 1854 & F1,8,L1 &   45.12 $\pm$   0.10        &  8.60 $\pm$ 0.26  \\ 
PG\,1309$+$355   & 0.184  &  4962 $\pm$  500 & 1       &   44.87 $\pm$   0.34        &  8.20 $\pm$ 0.37  \\ 
PG\,1351$+$640   & 0.087  &  2864 $\pm$  751 & F1,8,L1 &   44.65 $\pm$   0.06        &  7.57 $\pm$ 0.23  \\ 
PG\,1352$+$183   & 0.158  &  3927 $\pm$  395 & 1       &   44.91 $\pm$   0.09        &  8.03 $\pm$ 0.12  \\ 
PG\,1411$+$442   & 0.089  &  2617 $\pm$  260 & 1       &   44.46 $\pm$   0.08        &  7.36 $\pm$ 0.11  \\ 
PG\,1415$+$451   & 0.114  &  6768 $\pm$  680 & 1       &   44.36 $\pm$   0.13        &  8.11 $\pm$ 0.14  \\ 
PG\,1416$-$129   & 0.129  &  6944 $\pm$  415 & F1,8,L1 &   44.74 $\pm$   0.12        &  8.40 $\pm$ 0.12  \\ 
PG\,1426$+$015   & 0.086  &  4976 $\pm$  731 & F1,8,L1 &   45.21 $\pm$   0.10        &  8.44 $\pm$ 0.15  \\ 
PG\,1440$+$356   & 0.077  &  2913 $\pm$  295 & 1       &   44.75 $\pm$   0.07        &  7.65 $\pm$ 0.11  \\ 
PG\,1444$+$406   & 0.267  &  4226 $\pm$  425 & 12      &   45.58 $\pm$   (0.10)      &  8.56 $\pm$ 0.12  \\ 
PG\,1501$+$106   & 0.036  &  4673 $\pm$  470 & 1       &   44.22 $\pm$   0.09        &  7.69 $\pm$ 0.12  \\ 
PG\,1512$+$370   & 0.371  &  8333 $\pm$  835 & 1       &   45.48 $\pm$   0.07        &  9.08 $\pm$ 0.11  \\ 
PG\,1534$+$580   & 0.030  &  4987 $\pm$  500 & 1       &   43.86 $\pm$   0.10        &  7.50 $\pm$ 0.12  \\ 
PG\,1545$+$210   & 0.266  &  4796 $\pm$  480 & F13,L14 &   45.45 $\pm$   0.11        &  8.58 $\pm$ 0.13  \\ 
PG\,1613$+$658   & 0.129  &  8073 $\pm$~~~30 & F1,8,L1 &   45.09 $\pm$   0.07        &  8.78 $\pm$ 0.06  \\ 
PG\,1704$+$608   & 0.371  &  3894 $\pm$ 1469 & F1,8,L1 &   45.56 $\pm$   0.18        &  8.47 $\pm$ 0.37  \\ 
PG\,2130$+$099   & 0.061  &  3320 $\pm$  957 & F1,8,L1 &   44.54 $\pm$   0.08        &  7.62 $\pm$ 0.26  \\ 
PG\,2209$+$184   & 0.070  &  6595 $\pm$  660 & 1       &   44.41 $\pm$   0.17        &  8.13 $\pm$ 0.17  \\ 
PG\,2251$+$113   & 0.323  &  3758 $\pm$  375 & F8,L14  &   45.54 $\pm$   0.11        &  8.43 $\pm$ 0.13  \\ 
PG\,2308$+$098   & 0.432  &  5328 $\pm$  535 & 1       &   45.87 $\pm$   0.08        &  8.96 $\pm$ 0.11  \\ 
\enddata
\tablenotetext{a}{FWHM(\civ) measured in the single-epoch spectrum in units of km s$^{-1}$.}
\tablenotetext{b}{FWHM(\civ) and L$_{\lambda}$(1350\AA) are generally from the same reference,
	for the few exceptions: F: FWHM(\civ) reference, L: L$_{\lambda}$(1350\AA) reference.}
\tablenotetext{c}{$\log$\,[$\lambda L_{\lambda}$(1350\AA)/ergs s$^{-1}$]. Errors in () are 
	conservative errors assigned since no errors were quoted by the source study.} 
\tablenotetext{d}{The central mass (and uncertainties; see text) estimated based on single-epoch 
        UV spectroscopy, listed in logarithmic units.}
\tablerefs{
(1) Wang, Lu, \& Zhou 1998;
(2) O'Brien, \et 1998;
(3) Koratkar, \& Gaskell 1991;
(4) B.\ Peterson, 2001, private communication;
(5) Clavel, \et 1991;
(6) Korista, \et 1995;
(7) Wanders, \et 1997;
(8) Wilkes, \et 1999;
(9) This author's own measurements of data by Vestergaard \& Wilkes 2001;
(10) Laor, \et 1994;
(11) This author's own measurements of published profiles by Marziani, \et 1996;
(12) Laor, \et 1995;
(13) Wills, \et 1995;
(14) Schmidt, \& Green 1983; Kellerman, \et 1989.
}
\end{deluxetable}

\begin{deluxetable}{llccrr}
\tablewidth{0pt}
\tablecaption{Line Width and Luminosity Regression Parameters \label{regression}}
\tabletypesize{\small}
\tablehead{
\colhead{\bf Independent} &
\colhead{\bf Dependent} &
\colhead{\bf Sample} &
\colhead{\bf N} &
\colhead{\bf Slope} &
\colhead{\bf Intercept} \\
\colhead{\bf Variable} &
\colhead{\bf Variable} &
\colhead{} & 
\colhead{(\#)} &
\colhead{\bf $\pm$error} &
\colhead{\bf $\pm$error} 
}

\tablecolumns{6}
\startdata
 \multicolumn{6}{c}{\bf Bivariate Correlated Errors and Intrinsic Scatter Regression (BCES) } \\[3pt]
\tableline \\[-9pt]
Single-Epoch log L\tablenotemark{a} & Reverberation log L\tablenotemark{a} & A & 19 & 1.03$\pm$0.05 & $-$1.59$\pm$2.60 \\ 
\multicolumn{2}{l}{Bisector} & A & 19 &{\bf  1.04$\pm$0.05} &{\bf $-$1.93$\pm$2.35} \\ 
\tableline \\[-9pt]
Single-Epoch FW\tablenotemark{b} & Mean FW & B & 18 & 0.88$\pm$0.07 & 172$\pm$181 \\
\multicolumn{2}{l}{Bisector} & B & 18 & 0.87$\pm$0.07 & 206$\pm$187 \\ 
Single-Epoch FW & Mean FW & C & 17 & 0.96$\pm$0.10 & $-$33$\pm$251 \\
\multicolumn{2}{l}{Bisector} & C & 17 & {\bf 0.94$\pm$0.09} & {\bf 11$\pm$229} \\ 
\tableline \\[-9pt]
Single-Epoch FW & Rms FW & C & 17 & 0.97$\pm$0.09 & $-$237$\pm$266  \\
\multicolumn{2}{l}{Bisector} & C & 17 & {\bf 0.98$\pm$0.08} &{\bf  $-$265$\pm$227} \\ 
\enddata
\tablenotetext{a}{log [$\lambda L_{\lambda}$(5100\AA)/ ergs s$^{-1}$]} 
\tablenotetext{b}{FW = FWHM(\hb) ; Rms FW = FWHM(\hb, rms) ; Mean FW = FWHM(\hb, mean) } 
\tablecomments{
Sample A: All PG quasars in Kaspi \et (2000) \\
Sample B: Sample A except PG1351$+$640 (no reverberation FWHM(\hb)) \\
Sample C: Sample B except PG1613$+$658 (single-epoch FWHM(\hb) is not representative; see text) \\
}
\end{deluxetable}

\begin{deluxetable}{llccrr}  
\tablewidth{0pt}
\hspace{-2cm}
\tablecaption{Mass Regression Parameters $-$ Optical Measurements \label{optMregression}}
\vspace{-2cm}
\tabletypesize{\small}
\tablehead{
\colhead{\bf Independent} &
\colhead{\bf Dependent} &
\colhead{\bf Sam-} &
\colhead{\bf N} &
\colhead{\bf Slope} &
\colhead{\bf Intercept} \\
\colhead{\bf Variable} &
\colhead{\bf Variable} &
\colhead{\bf ple} & 
\colhead{(\#)} &
\colhead{\bf $\pm$error} &
\colhead{\bf $\pm$error} 
}

\tablecolumns{6}
\startdata
 \multicolumn{6}{c}{\bf Bivariate Correlated Errors and Intrinsic Scatter Regression (BCES) } \\[3pt]
\tableline \\[-9pt]
$\log$\,\mhbse\tablenotemark{a} & $\log$\,\mhbrms\tablenotemark{b} & C & 17 & {\bf 0.97$\pm$0.13} & {\bf 0.05$\pm$1.08}\\
  \multicolumn{2}{l}{BCES bisector}         & C & 17                &{\bf 1.03$\pm$0.14} & {\bf $-$0.49$\pm$1.17}\\
\tableline \tableline \\[-9pt]
$\log$\,\mhbse\ & $\log$\,\mhbmean\tablenotemark{b} & C & 17 & 0.97$\pm$0.08 & 0.11$\pm$0.68 \\
  \multicolumn{2}{l}{BCES bisector}          & C & 17 & 0.91$\pm$0.11 & 0.59$\pm$0.91 \\
\tableline \tableline \\[-9pt]
\multicolumn{6}{l}{Sample C: All PG quasars of Kaspi \et (2000) and Mrk\,110 and Mrk\,335 (Wandel } \\
\multicolumn{6}{l}{~~~\et 1999) but excluding PG1351$+$640 [no reverberation FWHM(\hb) is available]} \\
\multicolumn{6}{l}{~~~and PG1613$+$658 [single-epoch FWHM(\hb) is not representative; see text]} \\
\tableline 
\enddata
\tablenotetext{a}{\mhbse\ = Single-Epoch mass estimates based on \hb\ and optical continuum measurements.}
\tablenotetext{b}{$\log$\,\mhbrms{} and $\log$\,\mhbmean{} are the reverberation masses based on
	the \hb\ line widths only (in the rms and mean spectra, respectively) and the directly measured BLR 
	sizes in Table~6 by Kaspi \et (2000).}
\end{deluxetable}

\begin{deluxetable}{llcccr} 
\tablewidth{0pt}
\hspace{-2cm}
\tablecaption{Mass Regression Parameters $-$ UV Measurements \label{UVMregression}}
\tabletypesize{\small}
\tablehead{
\colhead{\bf Independent} &
\colhead{\bf Dependent} &
\colhead{\bf Sam-} &
\colhead{\bf N} &
\colhead{\bf Slope} &
\colhead{\bf Intercept} \\
\colhead{\bf Variable} &
\colhead{\bf Variable} &
\colhead{\bf ple} & 
\colhead{(\#)} &
\colhead{\bf $\pm$error} &
\colhead{\bf $\pm$error} 
}

\tablecolumns{6}
\startdata
 \multicolumn{6}{c}{\bf Bivariate Correlated Errors and Intrinsic Scatter Regression (BCES) } \\[3pt]
\tableline \\[-9pt] 
$\log$\,[FWHM$^{2}$(\civ) \lLl$^{0.7}$]\tablenotemark{a} & $\log$\,\mbh\tablenotemark{b} & 
		         UVrev & 26  & 0.40$\pm$0.12 & 7.16$\pm$0.21 \\
\multicolumn{2}{l}{BCES bisector}         & 
		         UVrev & 26  & 0.79$\pm$0.17 & 6.52$\pm$0.36 \\
\tableline \\[-9pt]
{\bf log\,[FWHM$^{\bf 2}$(C {\small IV}) $\lambda$L$_{\lambda}^{\bf 0.7}$]} &{\bf  log\,M$_{\rm \bf BH}$} 
                             & {\bf UVrev,b} & {\bf 25} & {\bf 0.54$\pm$0.19} & {\bf 6.91$\pm$0.33 }\\
\multicolumn{2}{l}{\bf BCES bisector}                    
	                     & {\bf UVrev,b} & {\bf 25} & {\bf 1.04$\pm$0.12} & {\bf 6.04$\pm$0.25} \\
\tableline \tableline \\[-9pt]
$\log$\,[FWHM$^{2}$(\civ) \lLl$^{0.7}$] & $\log$\,\mbh{}   & UV & 56  & 0.73$\pm$0.15 & 6.75$\pm$0.29 \\
\multicolumn{2}{l}{BCES bisector}                      & UV & 56  & 1.03$\pm$0.16 & 6.20$\pm$0.32 \\
\tableline \\[-9pt]
{\bf log\,[FWHM$^{\bf 2}$(C {\small IV}) $\lambda$L$_{\lambda}^{\bf 0.7}$]} &{\bf  log\,M$_{\rm \bf BH}$} &{\bf UVb } &{\bf 55} 
			&{\bf 0.91$\pm$0.12 }&{\bf 6.39$\pm$0.22} \\
\multicolumn{2}{l}{\bf BCES bisector}                    &{\bf UVb } & {\bf 55}  &{\bf 1.22$\pm$0.10} &{\bf 5.81$\pm$0.20} \\
\tableline \tableline \\[-9pt]
{\bf log\,M$_{\rm \bf BH, UV}$(C {\small IV}) } &{\bf  log\,M$_{\rm \bf BH}$ (bisector)} 
		&{\bf UVrev,b } &{\bf 25} &{\bf 1.07$\pm$0.14}&{\bf $-$0.61$\pm$1.11} \\
\tableline \\[-9pt]
{log\,M$_{\rm BH, UV}$(C {\small IV}) } &{log\,M$_{\rm BH}$ (bisector)} 
		&{ UVb } &{ 55} &{1.23$\pm$0.10 }&{$-$1.78$\pm$0.81} \\
\tableline \tableline \\[-9pt]
\multicolumn{6}{l}{Sample UVrev: All Seyfert~1s and PG quasars with reverberation masses from Wandel \et (1999) }\\
\multicolumn{6}{l}{~~~~~~~~and Kaspi \et (2000) with published FWHM(\civ) and UV continuum luminosity }\\
\multicolumn{6}{l}{Sample UVrev,b: Sample UVrev excluding NGC\,4151} \\
\multicolumn{6}{l}{Sample UV: Sample UVrev supplemented with PG quasars from BG92 with published FWHM(\civ) }\\
\multicolumn{6}{l}{~~~~~~~~and UV continuum luminosity } \\
\multicolumn{6}{l}{Sample UVb: Sample UV excluding NGC\,4151} \\
\enddata
\tablenotetext{a}{FWHM(\civ) is in units of 1000 \kms, and \lLl is in units of 10$^{44}$ ergs s$^{-1}$ and
	measured at 1350\,\AA.}
\tablenotetext{b}{For the objects with available reverberation mass determinations $\log$\,\mhbrms\ is used.
	For the additional PG quasars without such masses (sample UV and UVb only), the masses are estimated 
	using the single-epoch \mhb ; see section~\ref{optcal}}
\end{deluxetable}

\begin{deluxetable}{lccc}
\tablewidth{0pt}
\tablecaption{Probabilities of Mass Estimate Accuracies \label{probabilities.tab}}
\tabletypesize{\small}
\tablehead{
\colhead{\bf Calibration } &
\colhead{\bf Factor~3} &
\colhead{\bf Factor~6} &
\colhead{\bf Factor~10} \\
\colhead{\bf } &
\colhead{\bf (0.5 dex)} &
\colhead{\bf (0.78 dex)} & 
\colhead{\bf (1.0 dex)} \\
\colhead{\bf } &
\colhead{\bf accuracy} &
\colhead{\bf accuracy} &
\colhead{\bf accuracy} 
}

\tablecolumns{4}
\startdata
Optical & 14/18 $\approx$ 80\% & 16/18 $\approx$ 90\% & 17/18 $\approx$ 95\% \\[5pt]
\tableline \\[-5pt]
UV      & 18/26 $\approx$ 70\% & 22/26 $\approx$ 85\% & $\sim$23/26 $\approx$ 90\% \\[5pt]
\enddata
\end{deluxetable}

\clearpage


\setcounter{figure}{0}

\begin{figure}[]
\epsfxsize=7.0cm
\plottwo{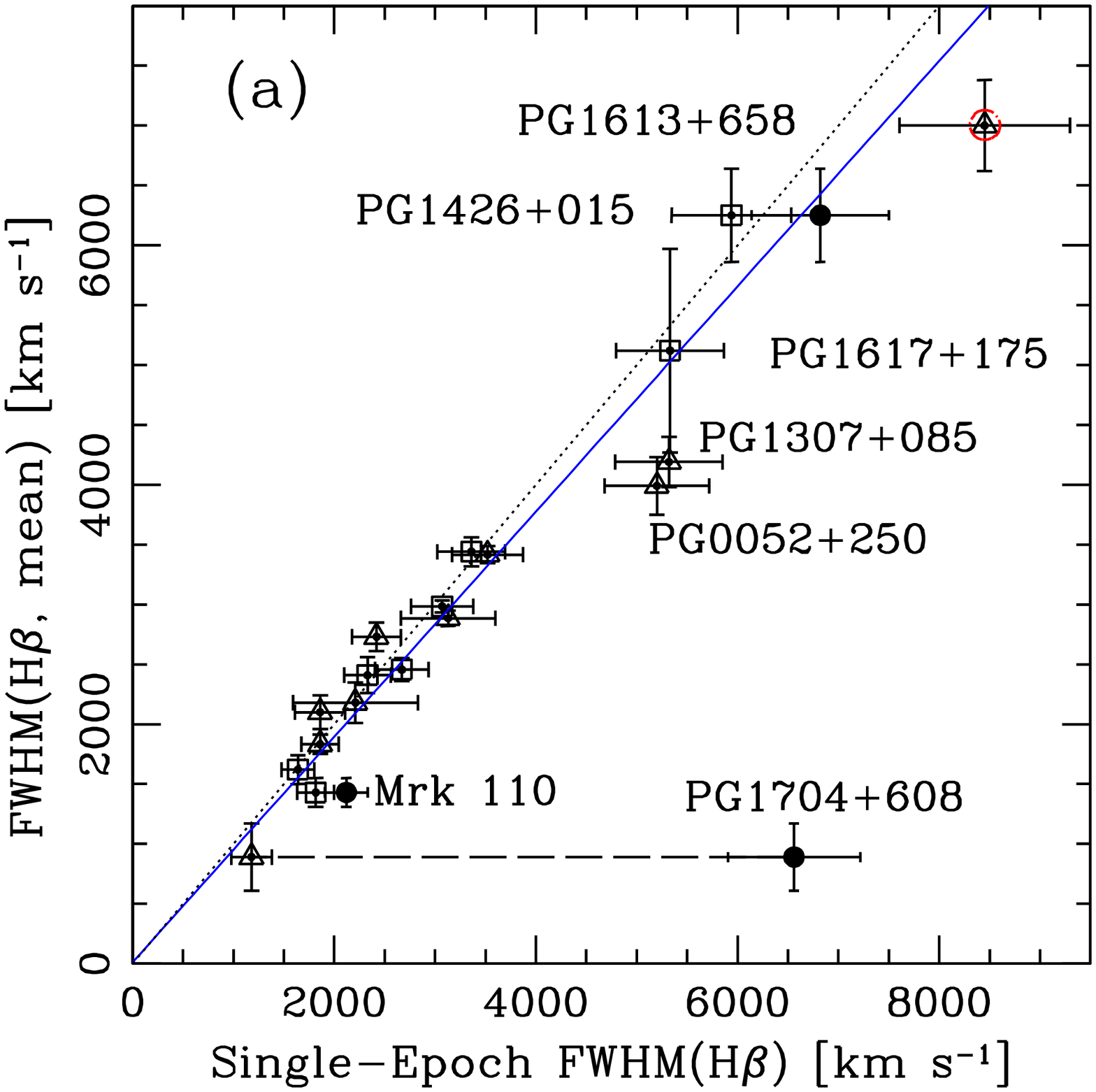}{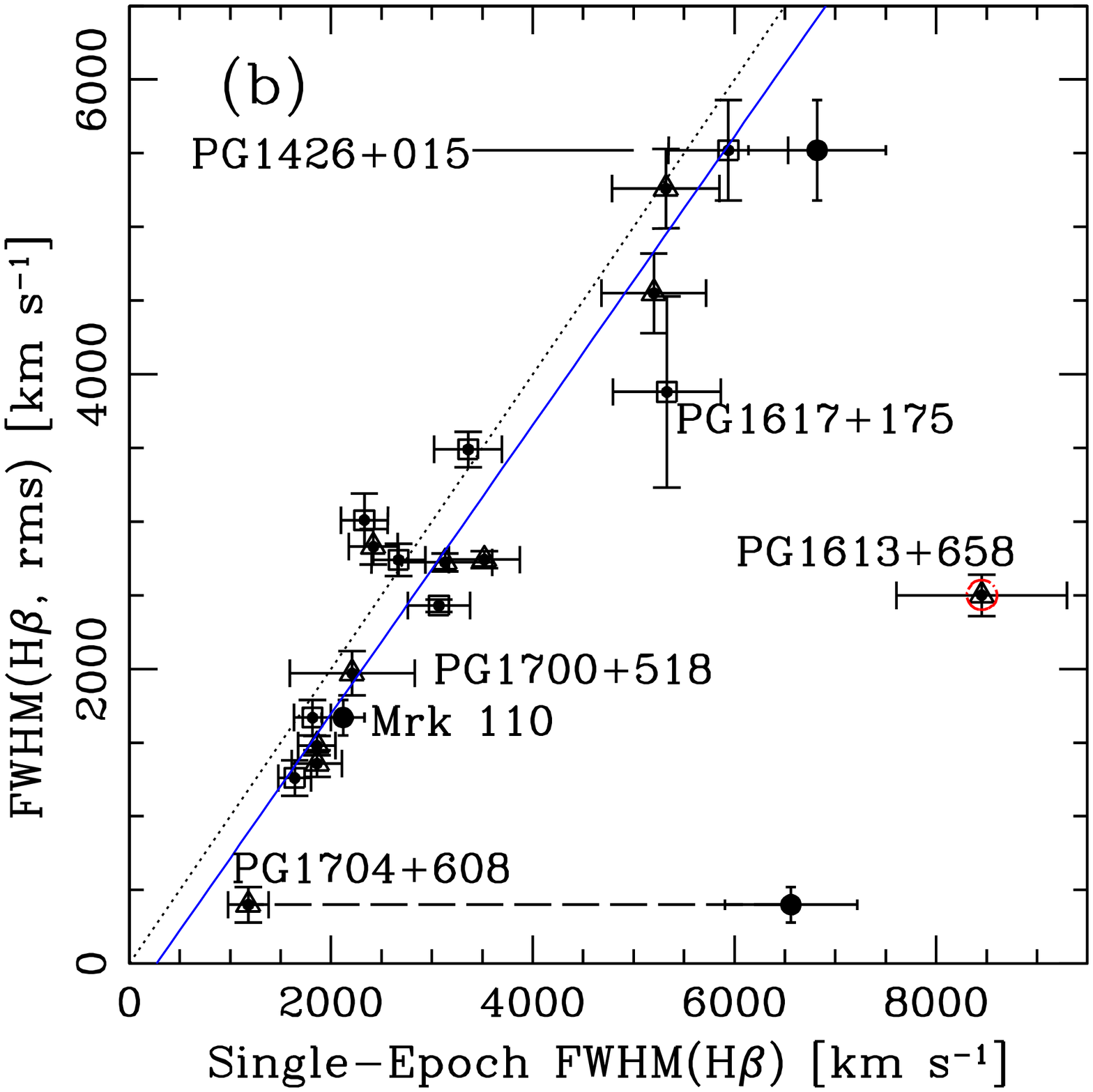}
\begin{center}
\caption[]{ The distribution of FWHM(\hb) in the mean (panel a) and rms
(panel b) multi-epoch spectra and single-epoch FWHM(\hb) of the PG quasars
presented by Kaspi \et (2000) and Mrk\,110 and Mrk\,335 (Wandel \et
1999).  The dotted line indicates a pure one-to-one relationship.
The objects discussed in the text are labeled. The open squares denote
Seyfert~1s while triangles show measurements for the quasars.
The three solid circles show the measurements of Boroson \& Green (1992) which
are based on spectra with both \feii{} emission and the narrow core component
subtracted. These FWHM measurements are not always representative of the BLR
velocity dispersion needed for this study (see text).
The solid line is the best fit BCES bisector regression line based on all
the objects in the diagram except PG1613$+$658 (sample C;
Table~\ref{regression}).  The BCES (Y$|$X) and (X$|$Y) regressions are not
plotted as they crowd the bisector.
The single-epoch FWHM(\hb) scatter around a one-to-one relationship with
FWHM(\hb, mean) to within 15\% $-$ 20\% variation and around a similar
relationship with FWHM(\hb, rms) to within 20\% $-$ 25\%.
Note, the ordinate range is different in the two diagrams.
} 
\end{center}
\end{figure}

\begin{figure}[h]
\vspace{1.0cm}
\epsfxsize=7.0cm
\plottwo{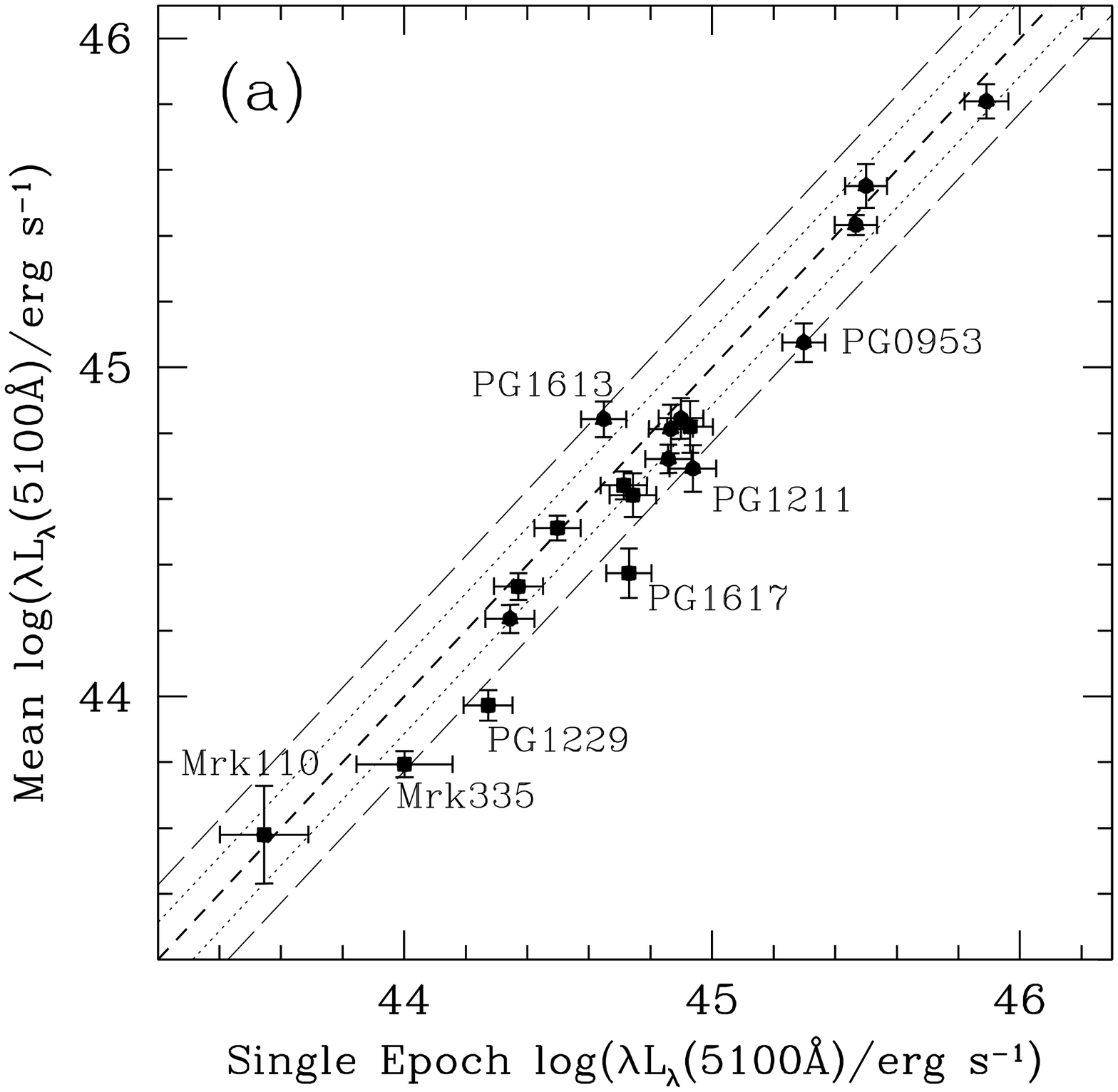}{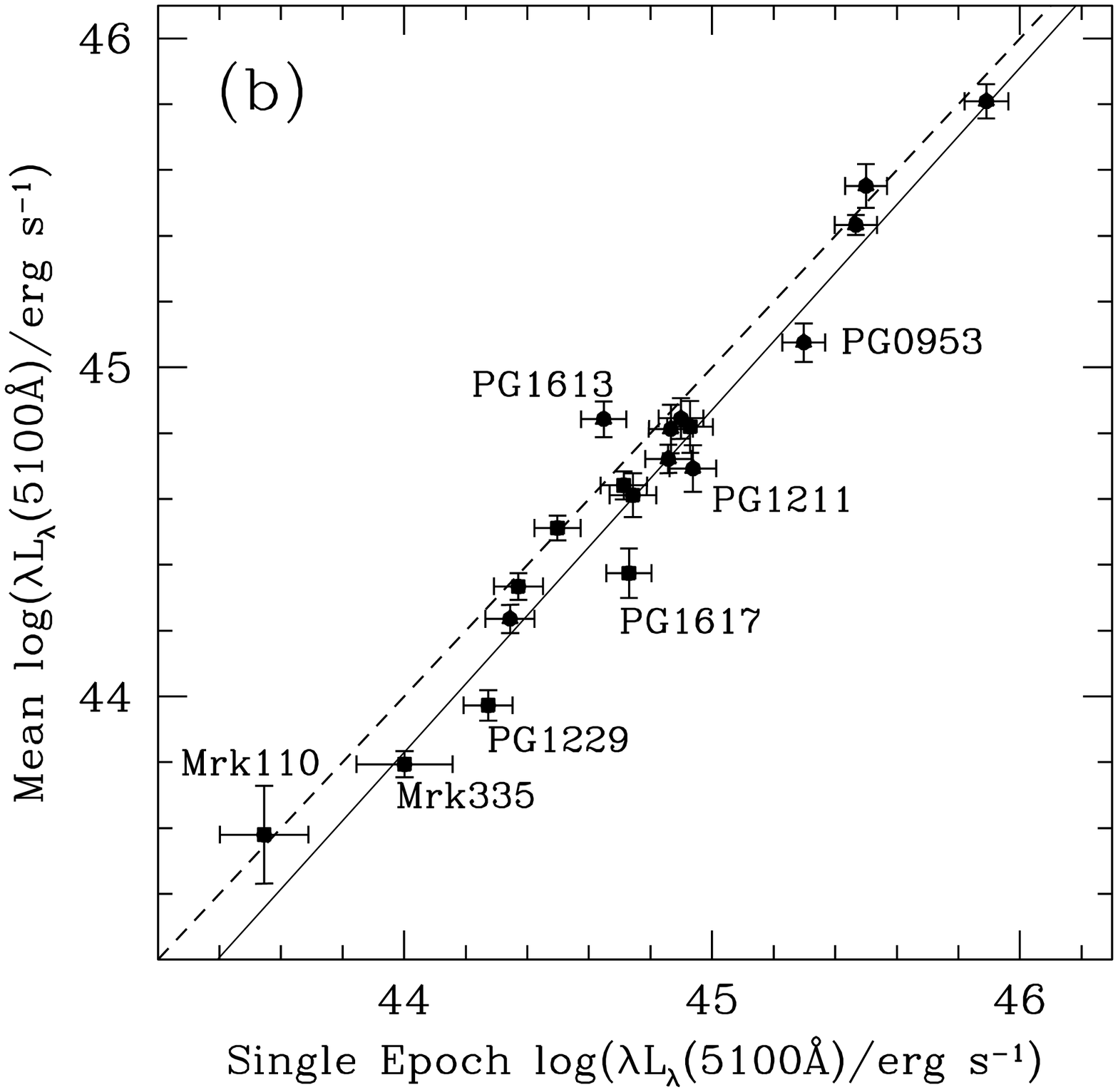}
\begin{center}
\caption[]{(a) The distribution of the mean $\lambda L_{\lambda}$(5100\AA)
multi-epoch measurements (Wandel \et 1999; Kaspi \et 2000) based on monitoring
data with
respect to the single-epoch $\lambda L_{\lambda}$(5100\AA) measurements
of Neugebauer \et (1987) and Schmidt \& Green (1983). The errors in the
mean $\lambda L_{\lambda}$(5100\AA) are the rms around this mean. The errors
in the single-epoch $\lambda L_{\lambda}$(5100\AA) are propagated errors
(see text). The short-dashed line (centrally positioned) denotes a one-to-one
relationship. The dotted and long-dashed lines represent $\pm$30\%
and $\pm$60\% luminosity variations, respectively (see text).
(b) The best fit regression lines are shown for the BCES bisector ({\it solid
line}). The BCES (Y$|$X) and (X$|$Y) regression lines would crowd the
bisector, if plotted. All these BCES regression fits are consistent with a
slope of 1.0. The single-epoch luminosities are offset by $+$0.138 dex at
$\lambda L_{\lambda}$(5100\AA) $\approx$\,44.8 ergs s$^{-1}$, the mid-range
luminosity, based on the BCES bisector.
} 
\end{center}
\end{figure}

\begin{figure}[ht]
\epsfxsize=7.0cm
\plottwo{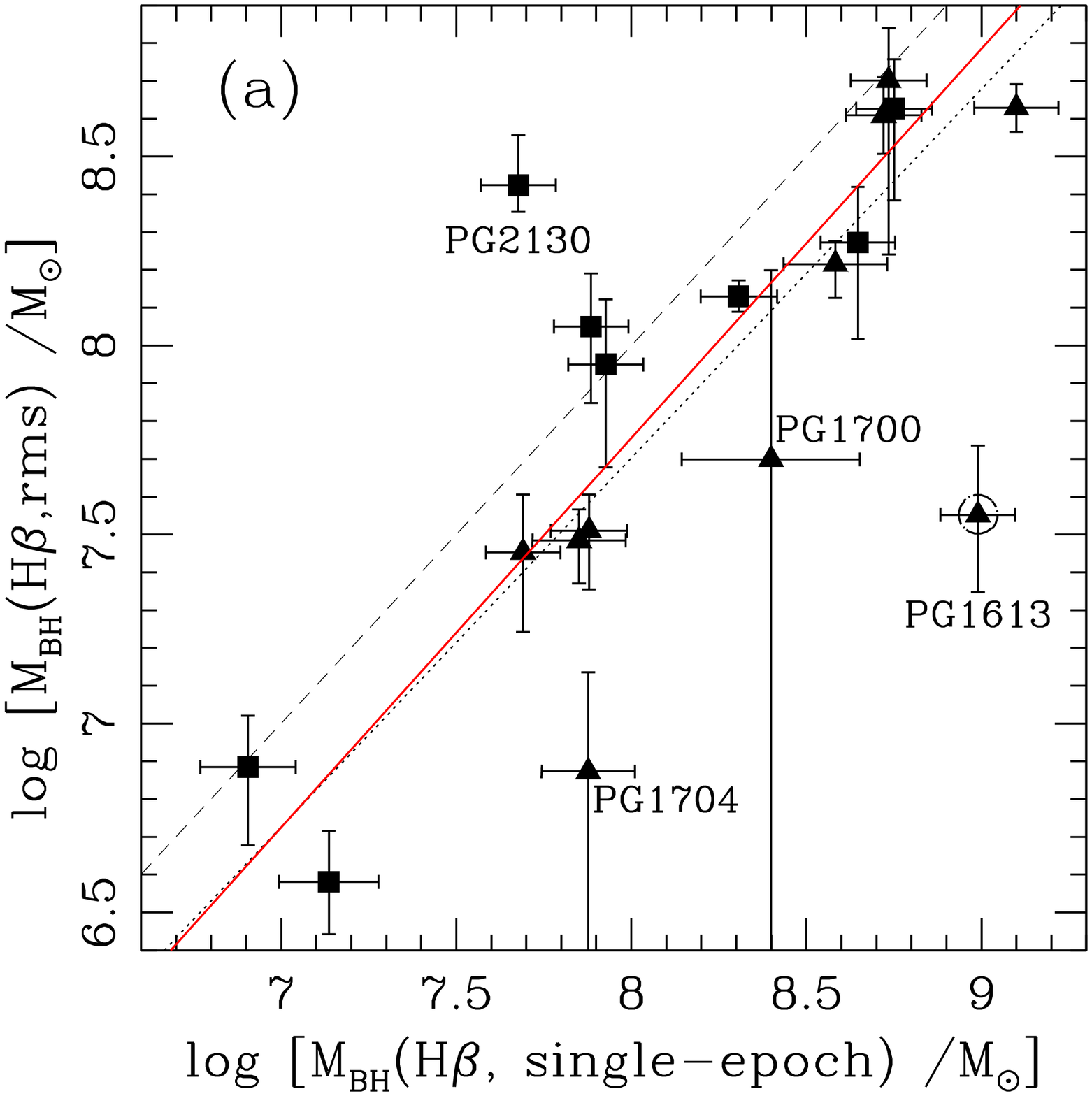}{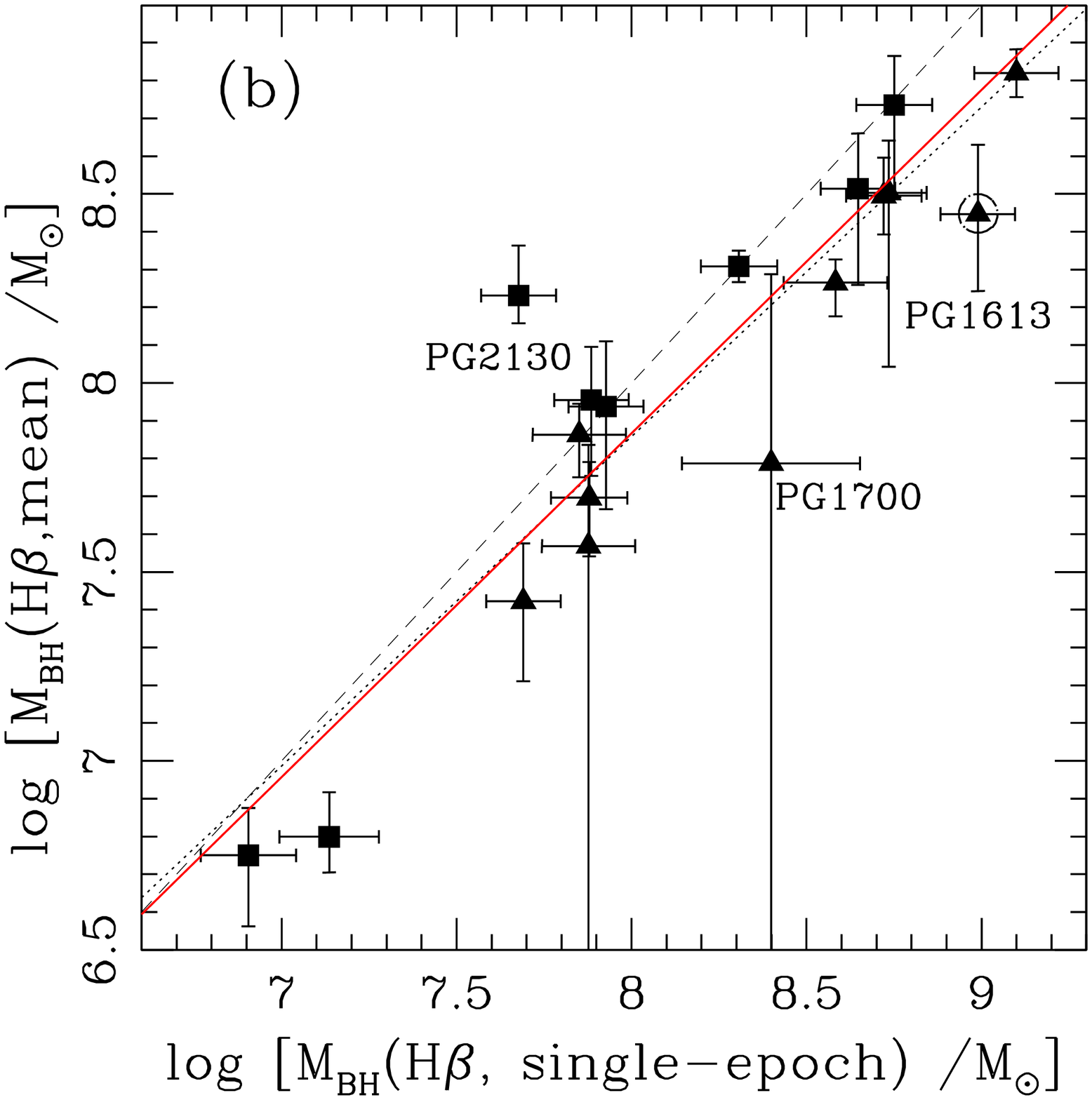}
\begin{center}
\caption[]{The reverberation masses derived from the rms (panel a) and
mean (panel b) spectra plotted versus the single-epoch mass estimates
based on optical spectral measurements.
Triangles denote quasars, while squares denote Seyfert\,1s. The dashed
line indicates a unity relationship. The dotted line is a BCES bisector
regression line to all the PG quasars with \hb{} measurements (shown),
while the solid line is the BCES bisector when PG1613 is excluded (see text).
As expected, \mhbse{} show less scatter with \mhbmean. Both relationships
are consistent with a one-to-one relationship to within the errors.
} 
\end{center}
\end{figure}

\begin{figure}[ht]
\epsfxsize=7.0cm
\plottwo{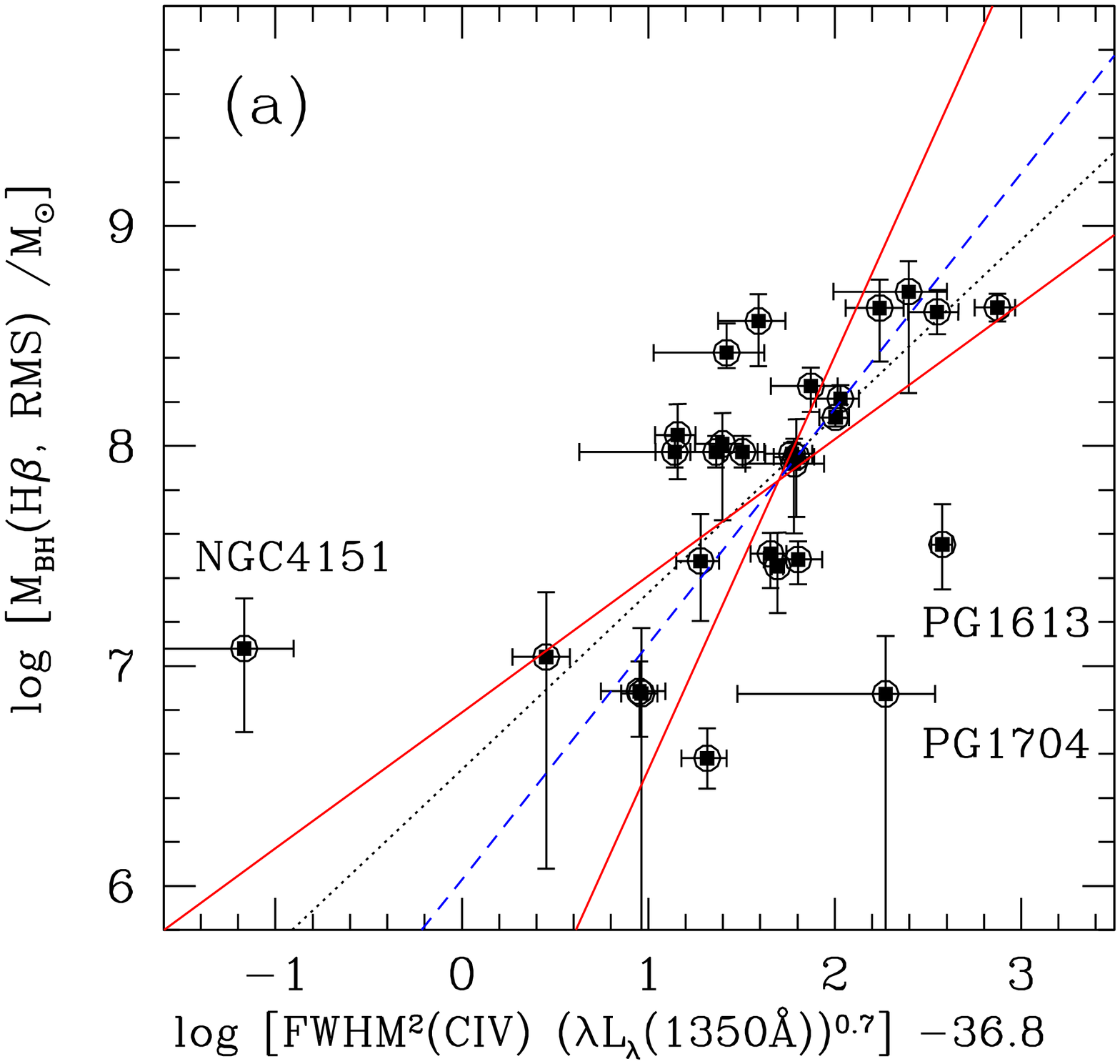}{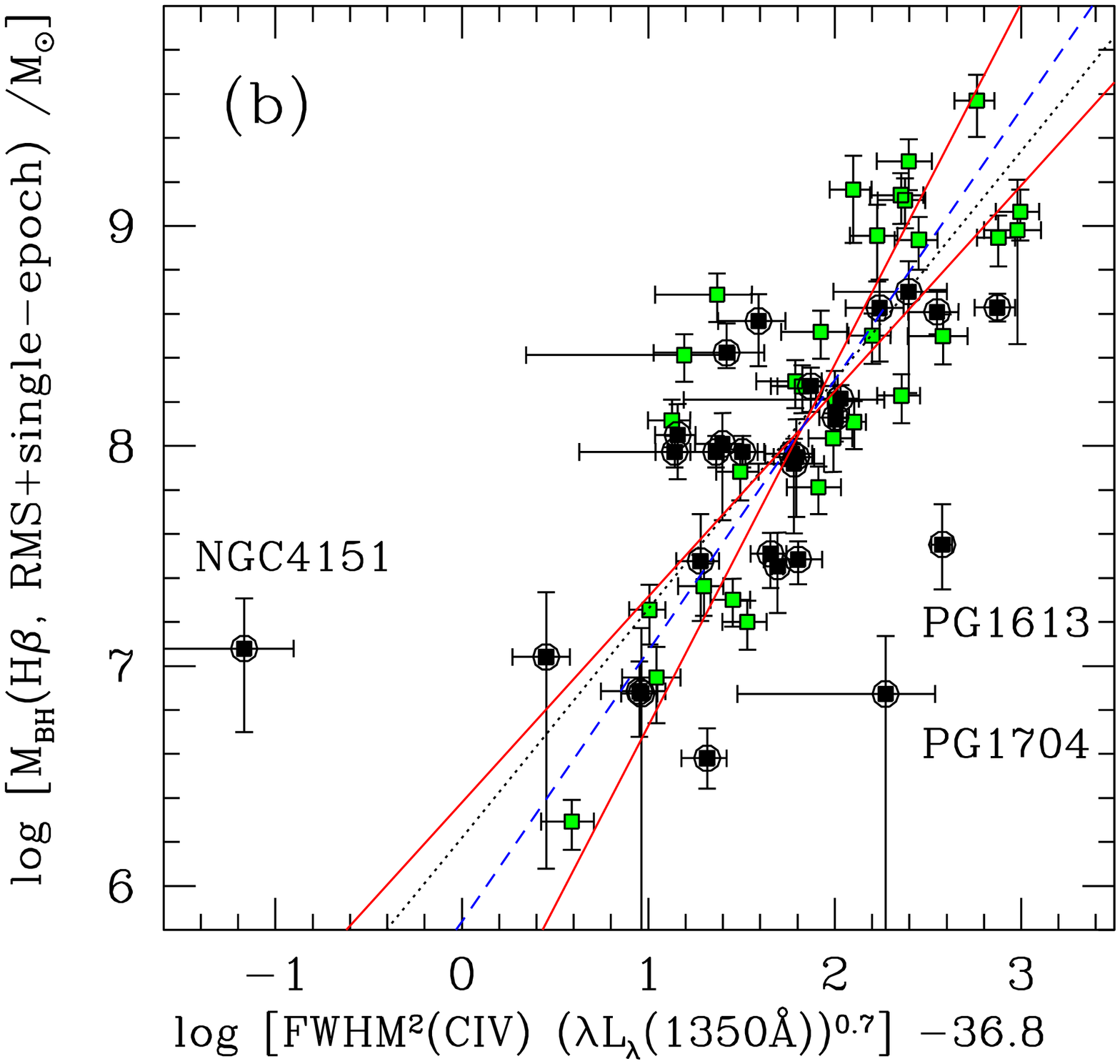}
\begin{center}
\caption[]{ Established and estimated central masses based on optical data
plotted versus the UV measurements. This distribution is the basis of the
calibration of the UV measurements. The ``optical masses'' (ordinate)
in panel (a) consist of the reverberation masses, \mhbrms, ({\it filled, 
circled squares}) derived from the rms spectrum and based on \hb{} only. 
In panel (b) these masses are supplemented with the single-epoch mass 
estimates, \mhbse, ({\it filled squares}) for the 30 PG quasars
with {\it no} reverberation mapping masses.
{\it Dotted lines:} BCES bisector regression lines to all the objects in each diagram.
{\it Solid lines:} BCES(Y$|$X) and (X$|$Y) regressions to all the objects except
NGC\,4151 (see text).
{\it Dashed lines:} the BCES bisector for all objects excluding NGC\,4151.
} 
\end{center}
\end{figure}

\begin{figure}[ht]
\epsfxsize=7.0cm
\plottwo{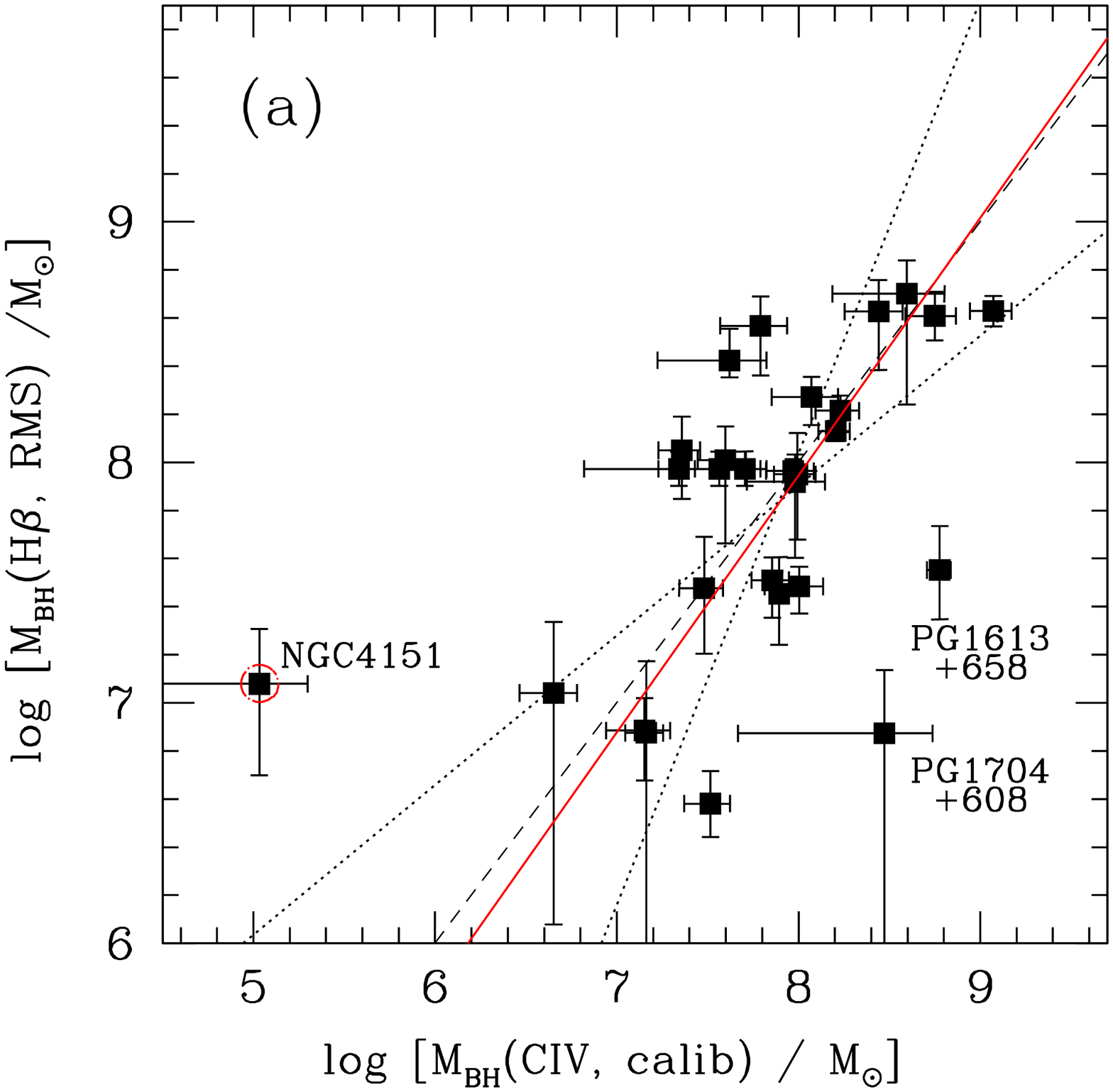}{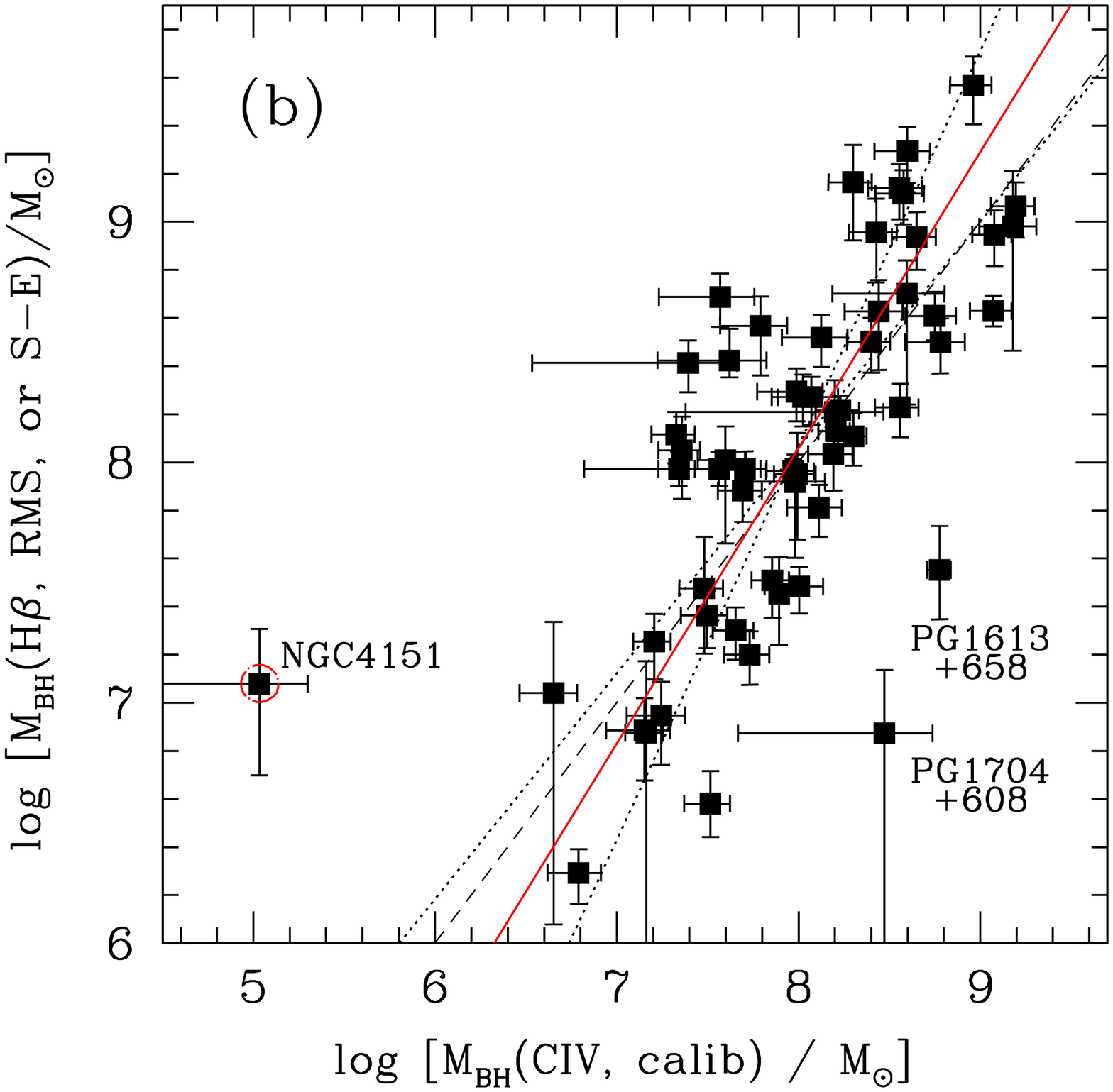}
\begin{center}
\caption[]{
The central mass estimates based on the calibration of the UV spectral
measurements are compared to central masses measured (and/or estimated;
panel b)
based on optical data.  {\it Dashed line:} pure unity relationship.
{\it Dotted lines:} BCES(Y$|$X) and BCES(X$|$Y) regression lines. {\it Solid line:}
the BCES bisector. NGC\,4151 was excluded from the regression analysis.
(a) \mbhuv{} estimates versus the established central masses, \mhbrms{}
[UVrev sample only].
(b) \mbhuv{} estimates are plotted for the full UV sample versus the
``optical masses'' described in Figure~\ref{Muv-Mrms.fig}b.
For both diagrams the mass relationships are consistent with
a unity relationship within the uncertainties.
} 
\end{center}
\end{figure}

\begin{figure}[ht]
\epsfxsize=7.0cm
\plottwo{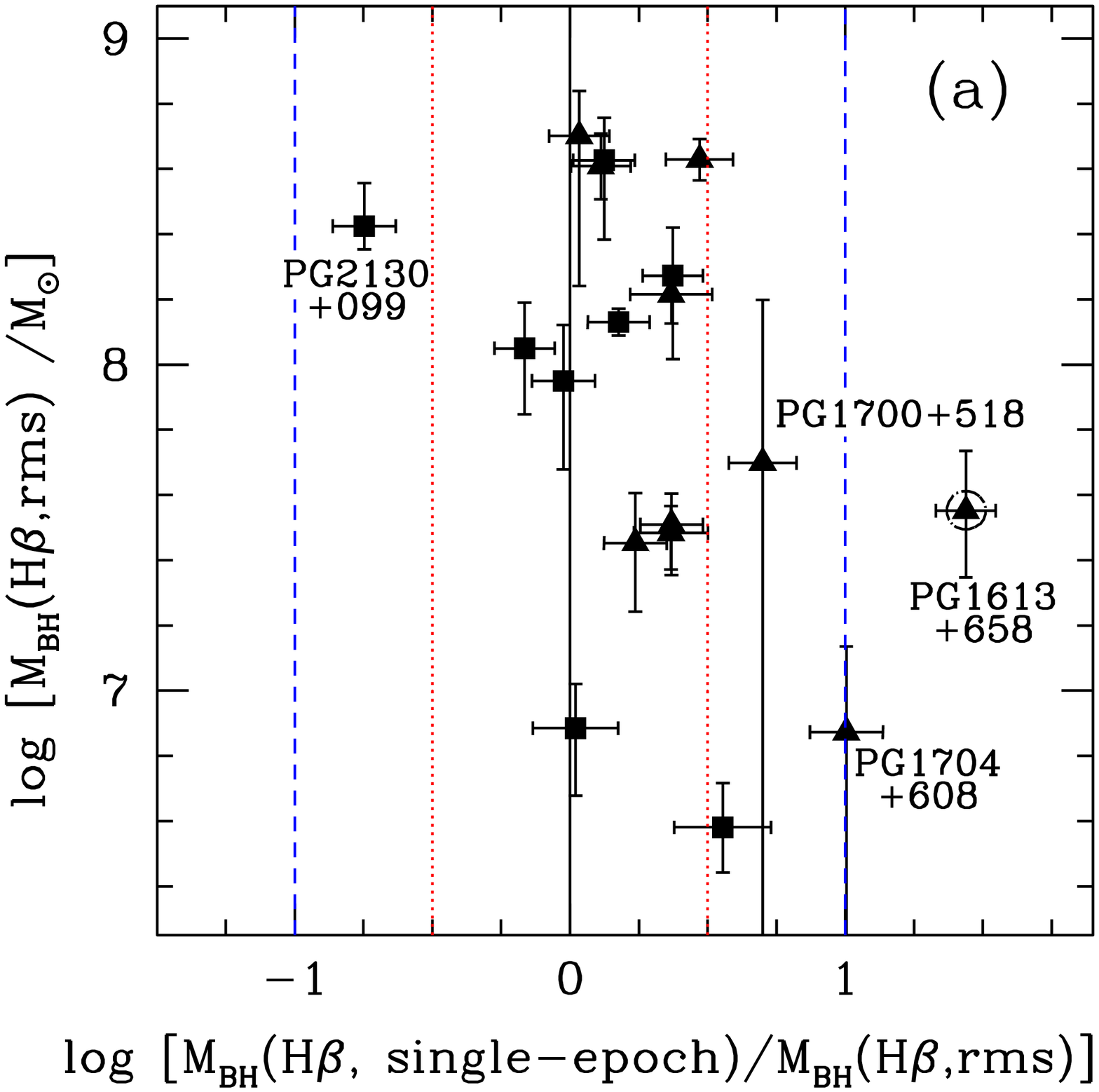}{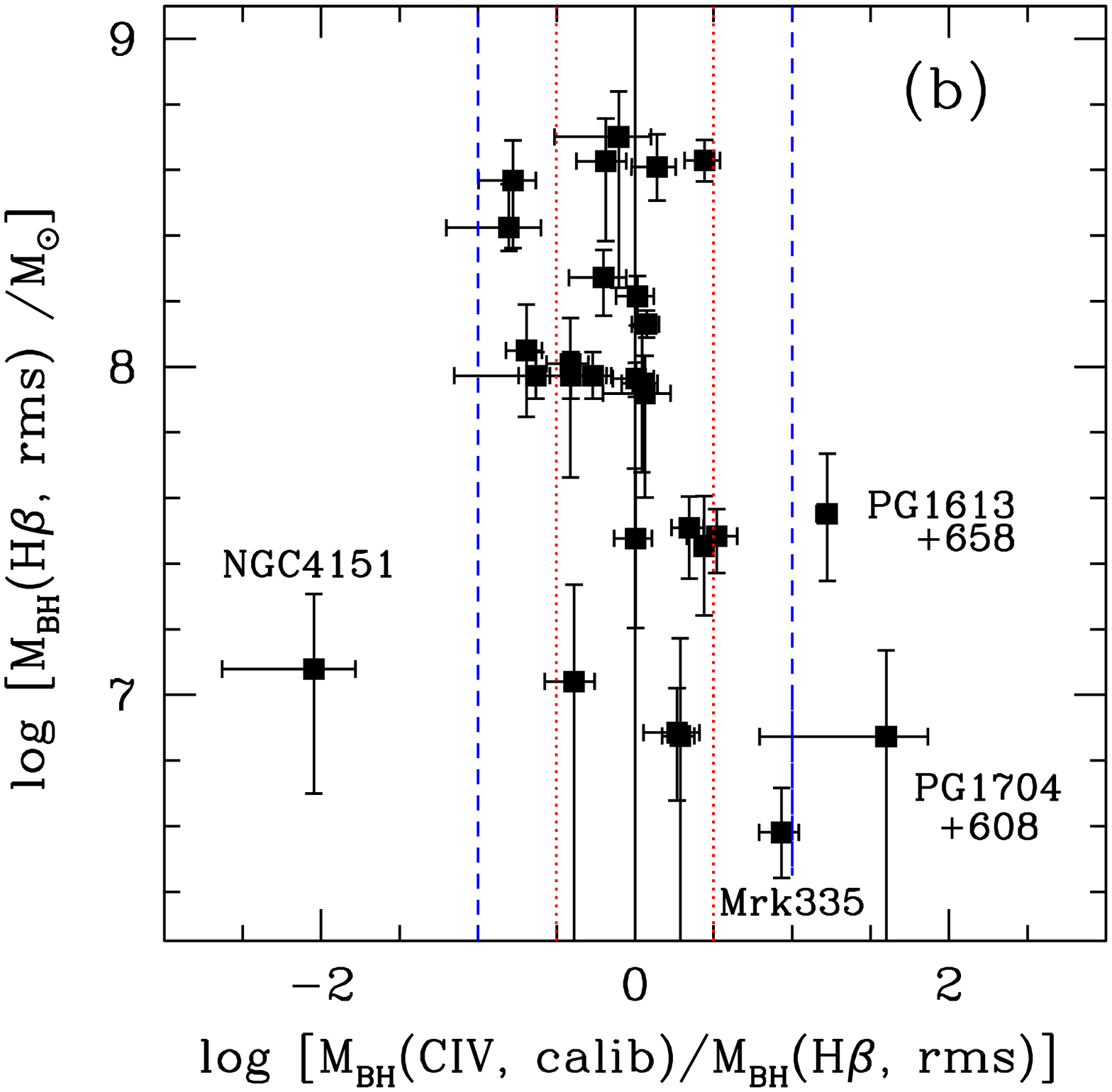}
\begin{center}
\caption[]{The established central masses, \mhbrms, based on optical multi-epoch
spectral measurements plotted versus the deviations of mass estimates based on
calibrated single-epoch spectra.  (a) Deviations of masses
estimated from optical spectra, i.e., \mhbse\  divided by \mhbrms. 
(b) Deviations in the masses based on UV spectral
measurements, \mbhuv(\civ). 
The uncertainties in the abscissa are the (propagated) uncertainties in
the single-epoch masses (\ie {\it not} the mass deviation error).
A strictly unity relationship is indicated by the solid line. Offsets of
$\pm$0.5\,dex ($\pm$1\,dex) are indicated by the dotted (dashed) lines.
} 
\end{center}
\end{figure}

\begin{figure}[ht]
\epsfxsize=7.0cm
\plottwo{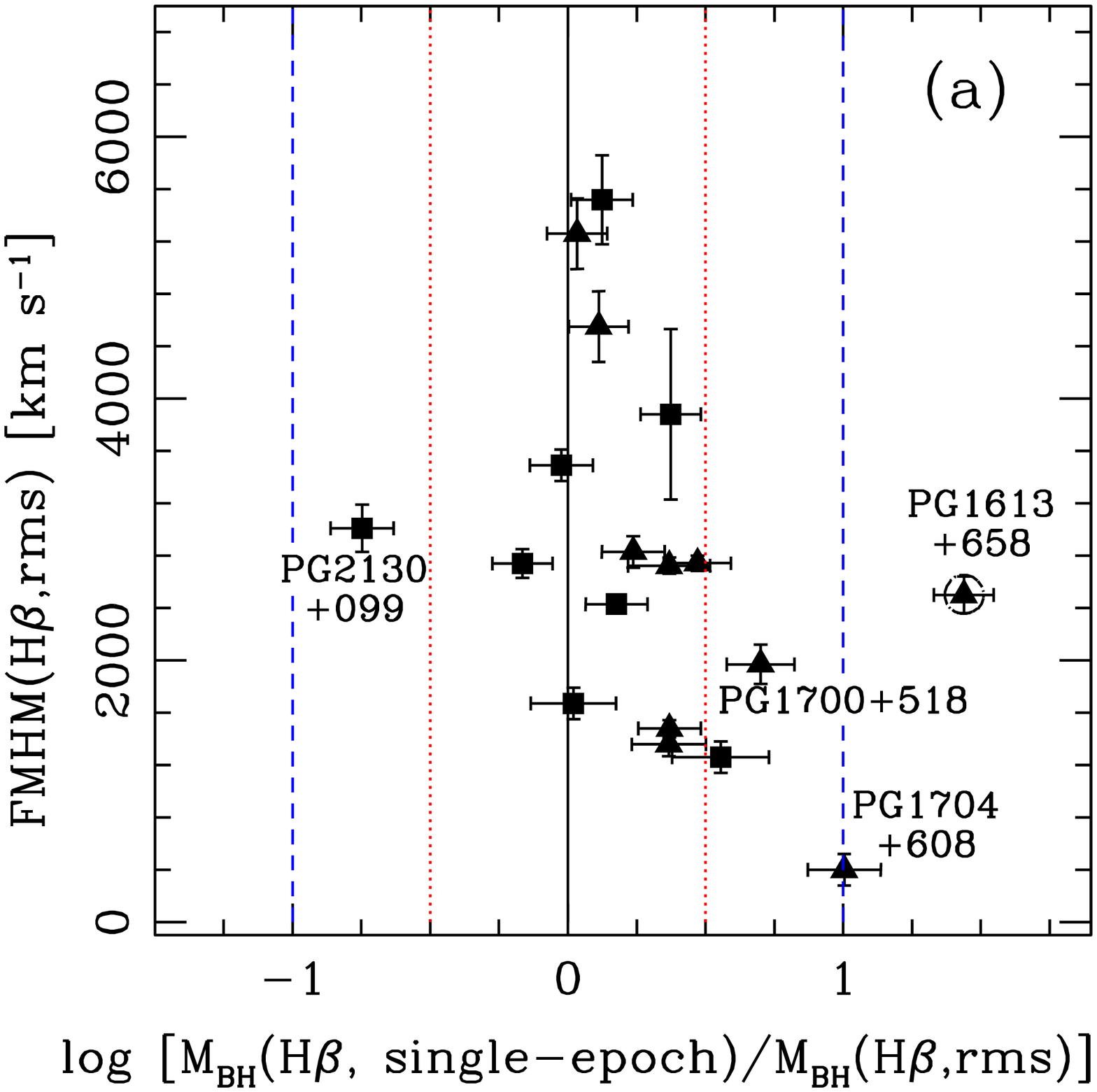}{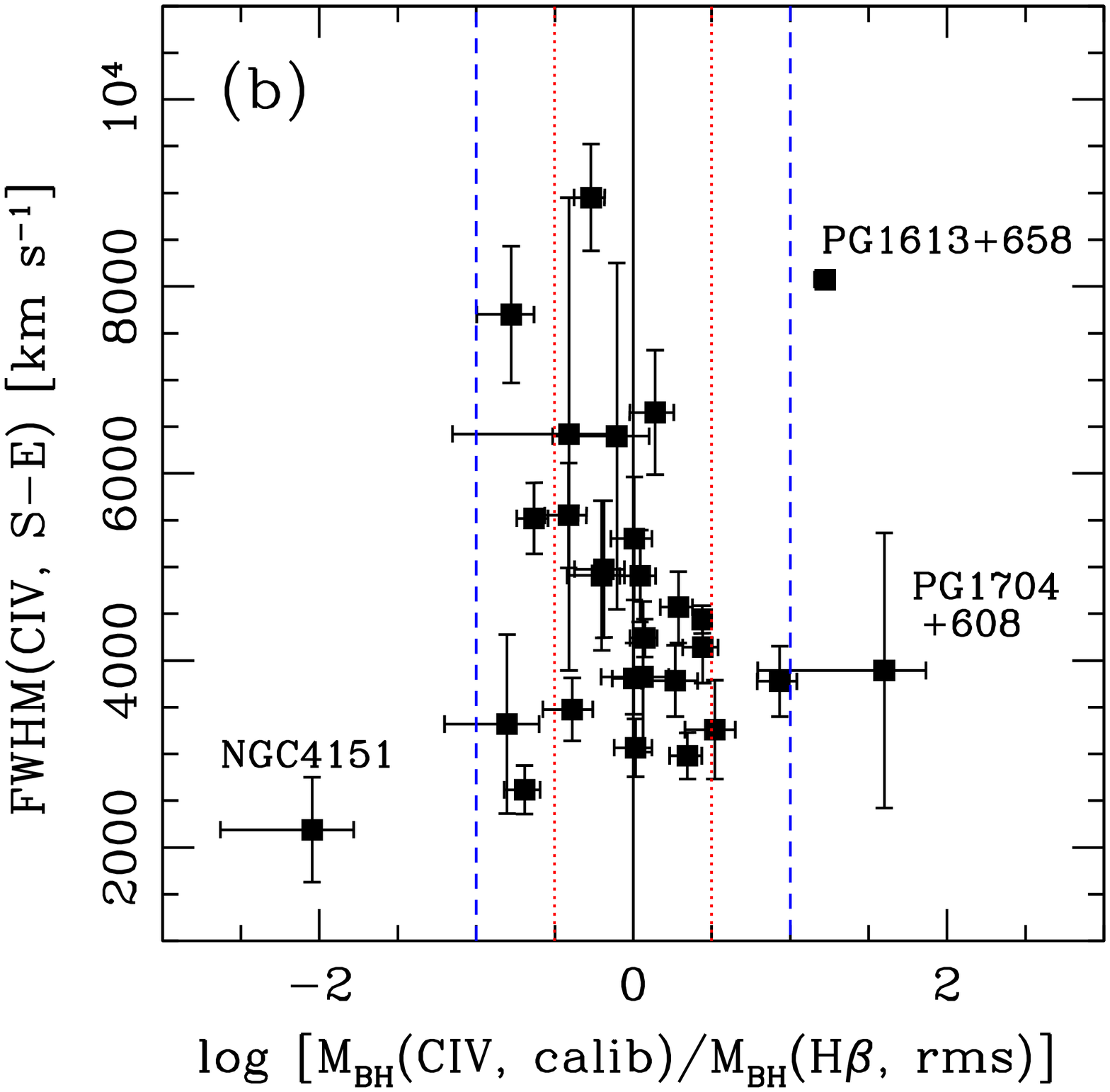}
\begin{center}
\caption[]{The mass deviations from Figure~\ref{MCal-problty.fig} plotted here
versus the \hb{} and \civ{} line widths. The optical mass deviations are plotted
against FWHM(\hb, rms), the width of the rms profile (i.e., the variable part),
in (a) and versus the single-epoch FWHM(\civ) in (b).
Lines and symbols are as in Figure~\ref{MCal-problty.fig}.
Notice that the larger mass discrepancies tend to occur for the most narrow
lined objects in both cases (or those with the strongest variability occurring
in the narrow line core: PG1613$+$658 and PG1704$+$608).
} 
\end{center}
\end{figure}

\clearpage

\end{document}